\documentclass[10pt,twocolumn,letterpaper]{article}

\usepackage[switch]{lineno}
\usepackage{iccv}
\usepackage{times}
\usepackage{epsfig}
\usepackage{graphicx}
\usepackage{amsmath}
\usepackage{amssymb}
\usepackage{algorithm}
\usepackage{algorithmic}
\usepackage{bbm}
\usepackage{amsfonts}
\usepackage{mathrsfs}
\usepackage{multirow}

\usepackage{microtype}
\usepackage{graphicx}
\usepackage{subcaption}
\usepackage{booktabs} 
\usepackage[flushleft]{threeparttable}
\usepackage{setspace}
\usepackage{multirow}
\usepackage{array}
\usepackage{silence}
\usepackage{fix-cm}
\newcommand{\PreserveBackslash}[1]{\let\temp=\\#1\let\\=\temp}
\newcolumntype{C}[1]{>{\PreserveBackslash\centering}p{#1}}
\newcolumntype{R}[1]{>{\PreserveBackslash\raggedleft}p{#1}}
\newcolumntype{L}[1]{>{\PreserveBackslash\raggedright}p{#1}}
\usepackage{xcolor}
\usepackage{ulem}
\usepackage{soul}
\usepackage{makecell}
\usepackage{cancel}
\setulcolor{blue}
\setul{0.5pt}{0.6pt}

\makeatletter
\newcommand\notsotiny{\@setfontsize\notsotiny{6.2}{7}}
\makeatother

\usepackage{amsmath}
\usepackage{amssymb}
\usepackage{mathtools}
\usepackage{amsthm}
\theoremstyle{plain}

\theoremstyle{definition}

\theoremstyle{remark}

\usepackage[textsize=tiny]{todonotes}

\usepackage[pagebackref=true,breaklinks=true,letterpaper=true,colorlinks,bookmarks=false]{hyperref}

\iccvfinalcopy

\usepackage[capitalize,noabbrev]{cleveref}

\begin{document}

\title{
Multi-Task Fine-Tuning \\Enables Robust Out-of-Distribution Generalization in Atomistic Models
}

\author{
      Chengqian Zhang{$^{1,2}$},~
      Duo Zhang{$^{1,2}$},~
      Anyang Peng{$^{1}$},~
      Mingyu Guo{$^{3,4}$},~
      Yuzhi Zhang{$^{3}$},~\\
      Lei Wang{$^{5}$},~
      Guolin Ke{$^{3}$},~
      Linfeng Zhang{$^{1,3}$},~
      Tiejun Li{$^{2,6,7\dag}$},~
      Han Wang{$^{8,9\dag}$}
      \\\vspace{-8pt}{\small~}\\
    \small {$^{1}$}AI for Science Institute, Beijing, China;\\
    \small {$^{2}$}Center for Data Science, Peking University, Beijing, China;\\
    \small {$^{3}$}DP Technology, Beijing, China;\\
    \small {$^{4}$}School of Chemistry, Sun Yat-sen University, Guangzhou, China;\\
    \small {$^{5}$}Beijing National Laboratory for Condensed Matter Physics,\\
    \small Institute of Physics, Chinese Academy of Sciences, Beijing, China;\\
    \small {$^{6}$}LMAM and School of Mathematical Sciences, Peking University, Beijing, China;\\
    \small {$^{7}$}Center for Machine Learning Research, Peking University, Beijing, China;\\
    \small {$^{8}$}National Key Laboratory of Computational Physics,\\
    \small Institute of Applied Physics and Computational Mathematics, Beijing, China;\\
    \small {$^{9}$}HEDPS, CAPT, College of Engineering, Peking University, Beijing, China;\\
    \small{$^{\dag}$}Corresponding to: \tt{tieli@pku.edu.cn} \tt{wang\_han@iapcm.ac.cn}, 
}

\maketitle

\begin{abstract}

Accurate \textit{de novo} molecular and materials design requires structure–property models that generalize beyond known regimes.
Although pretrained atomistic models achieve strong in-distribution accuracy after fine-tuning, their reliability under out-of-distribution (OOD) conditions remains unclear.
We identify a critical failure mode in downstream adaptation: standard fine-tuning induces representation collapse, erasing pretrained chemical and structural priors and severely degrading OOD performance. 
To address this limitation, we propose multi-task fine-tuning (MFT), which jointly optimizes downstream property prediction with a physically grounded force-field objective inherited from pretraining.
This approach preserves essential chemical priors while enabling task-specific adaptation. 
Across molecular and materials benchmarks, MFT consistently improves OOD generalization, approaching the theoretical limit set by in-distribution accuracy, while outperforming standard fine-tuning, training from scratch, and state-of-the-art task-specific models.
These results establish safe adaptation as a central requirement for large atomistic models and position MFT as a practical and data-efficient pathway toward robust molecular and materials discovery.

\end{abstract}

\section{Introduction}

\textit{De novo} design of molecules and materials with exceptional properties remains highly challenging~\cite{sanchez2018inverse}.
A crucial component of this endeavor is accurate modeling of structure–property relationships. 
These relationships enable two complementary strategies: (i) high-throughput virtual screening of databases to identify candidates with target properties~\cite{curtarolo2013high}, and (ii) conditional generative modeling to produce novel structures that meet specified property constraints~\cite{cao2025space, bao2023equivariant}. 
Atomistic machine learning (ML)  models that learn mappings from three-dimensional atomic structures (coordinates and species) to properties consistently achieve superior predictive accuracy compared with models based solely on chemical formulae, 1D string representations (e.g., SMILES), or 2D planar graph representations~\cite{zhou2023unimol,antoniuk2025boom}.
Consequently, atomistic ML models for property prediction have become indispensable tools in molecular and materials design.

\textit{De novo} design aims to discover molecules and materials with properties, chemical compositions, or configurational features beyond those documented to date, and therefore requires models that can reliably generalize to out-of-distribution (OOD) molecules and materials.
However,
recent benchmark studies have highlighted a substantial degradation in OOD generalization, manifested as a pronounced performance gap between in-distribution (ID; drawn from the training data distribution) and OOD tasks, for ML models in property prediction.
In the domain of organic molecules, the BOOM benchmark~\cite{antoniuk2025boom} defines OOD splits based on molecules whose target property values lie significantly above or below the range covered by the training data, and reveals that even top-performing models exhibit average OOD errors up to 3 times higher than their ID performance.
Shimakawa et al.~\cite{shimakawa2024extrapolative} further demonstrate severe performance degradation of ML property-prediction models under both property-based and structure-based OOD settings, particularly for tasks with limited training data.
For inorganic crystalline materials, OOD benchmarks, either identifying lowest-density regions in feature or property space~\cite{sadman2024oodmatbench} or applying leave-one-out protocols in composition or clustered feature space~\cite{sadman2024oodmatbench, li2025probing}, 
consistently demonstrate that state-of-the-art (SOTA) graph neural networks (GNNs) exhibit substantial accuracy degradation when evaluated outside their training domain.
Collectively, these findings underscore a critical and largely unresolved challenge: severe degradation in OOD generalization is an almost universally observed phenomenon in atomistic property prediction, across both molecular and materials systems, and largely independent of the specific definition of the OOD regime.

Pretraining on large-scale datasets to encode prior chemical knowledge into the representation (feature) space, followed by fine-tuning for downstream property prediction tasks, has proven effective in improving model generalizability.
Self-supervised pretraining strategies, such as coordinate denoising and atom-type masking, have been shown to learn robust representations that significantly enhance downstream property prediction performance~\cite{zaidi2023pretraining, zhou2023unimol, pmlr-v202-shu23a}.
Similarly, supervised pretraining on force-field tasks~\cite{batatia2025foundation, kim2024data, rhodes2025orb, yang2024mattersim, wood2025uma, bochkarev2024graph}, 
namely, training on datasets of molecular or materials configurations labeled by first-principles calculations such as density functional theory (DFT)~\cite{barroso2024open, chanussot2021open, levine2025open}, has been shown to encode chemically intuitive representations~\cite{zhang2024dpa}, leading to substantial accuracy gains upon fine-tuning for materials and molecular property prediction~\cite{shoghi2023molecules, yang2024mattersim, kong2025mattertune}.
However, the performance gains achieved by leveraging such prior chemical knowledge are almost exclusively evaluated under the ID settings.
The fundamental question of whether fine-tuning pretrained models for property prediction tasks can reliably improve OOD performance remains largely underexplored.

Pretrained vision and language models often exhibit strong zero-shot performance~\cite{chen2020moco, guo2025deepseek}, whereas task-specific fine-tuning can be detrimental to OOD generalization through \textit{representation collapse}, the distortion of general-purpose prior knowledge during downstream adaptation~\cite{aghajanyan2021better, pmlr-v139-radford21a}.
In contrast, preserving pretrained representations such as through linear probing, yields superior OOD robustness, a gap widening as the quality of the pretrained representations improves~\cite{ananya2022distort}.

\begin{figure*}[t]
  \begin{center}
    \centerline{\includegraphics[width=2\columnwidth]{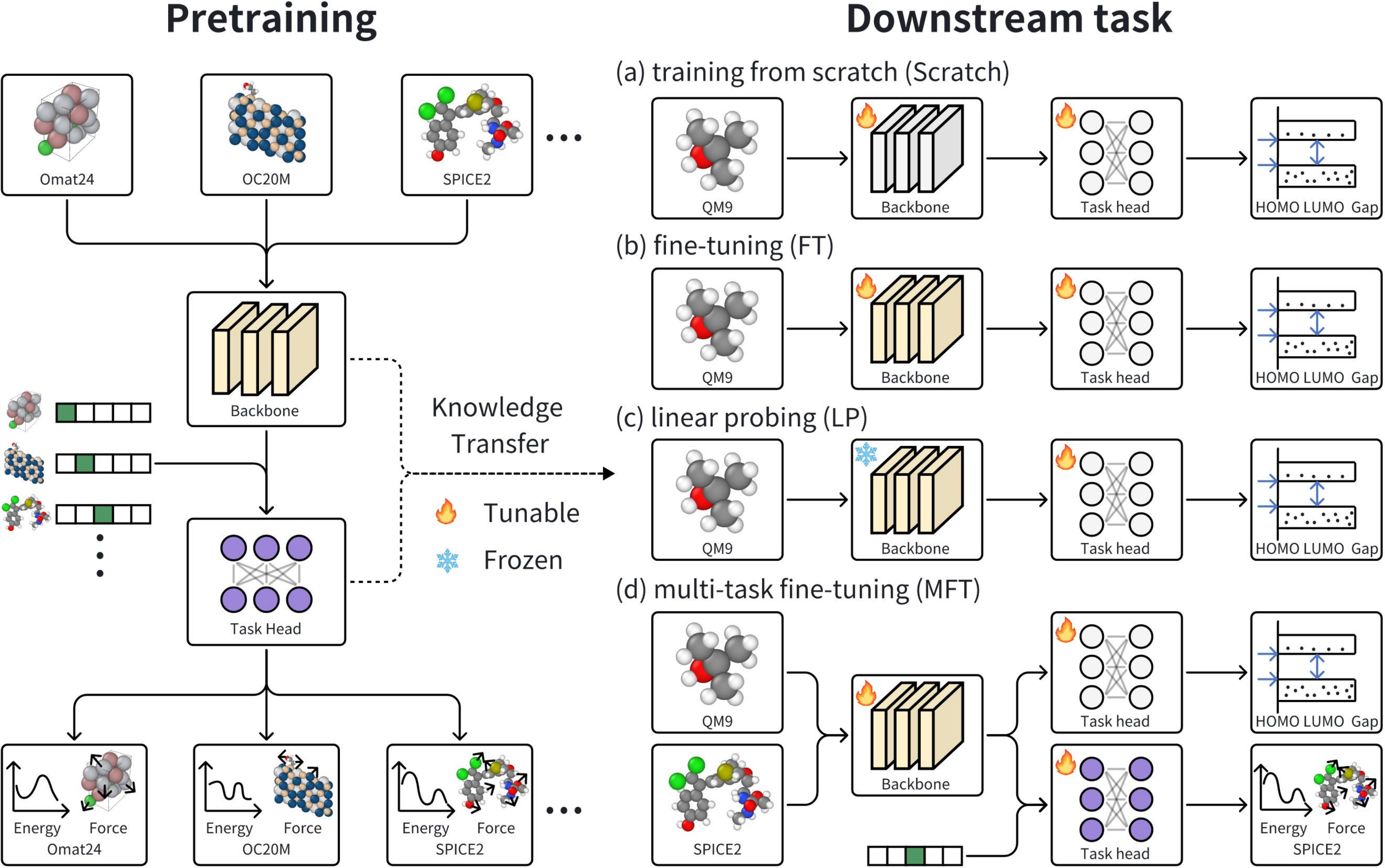}}
    \caption{Schematic illustration of pretraining and downstream adaptation strategies.
    The left panel depicts the pretraining phase.
    The right panel compares four downstream adaptation strategies:
    (a) training from scratch (Scratch), in which both the backbone and the task-specific prediction head are randomly initialized and optimized end-to-end;
    (b) fine-tuning (FT), where the backbone is initialized from the pretrained model and all parameters are jointly updated;
    (c) linear probing (LP), which freezes the pretrained backbone and optimizes only a randomly initialized task-specific head; and
    (d) multi-task fine-tuning (MFT), which jointly optimizes the pretrained backbone together with two heads: a task-specific head for property prediction and an auxiliary force-field head inherited from the pretraining stage.
    }
    \label{fig:overview}
  \end{center}
  \vskip -0.3in
\end{figure*}

\begin{figure*}[t]
  \begin{center}
    \centerline{\includegraphics[width=2\columnwidth]{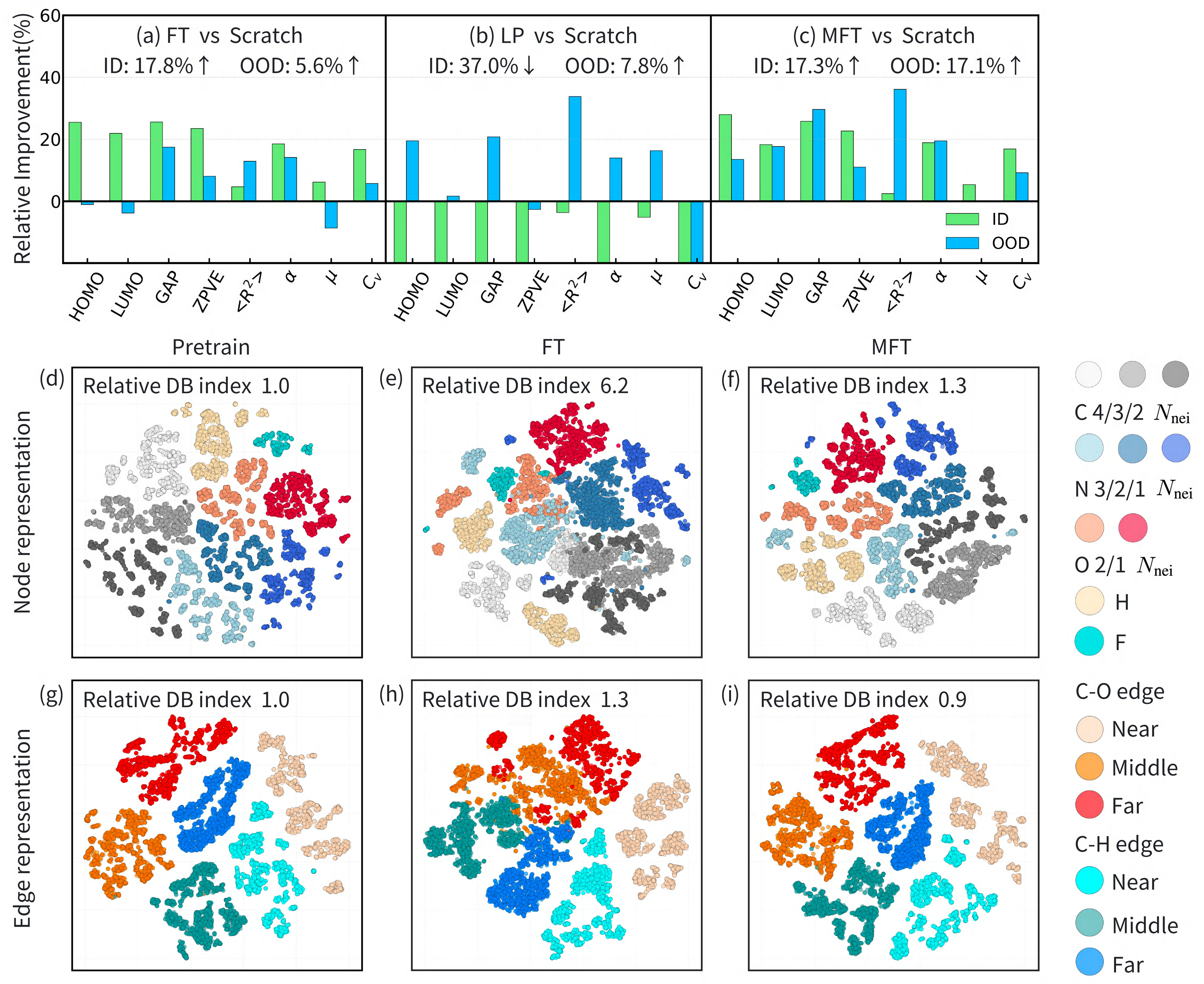}}
    \caption{
    Analysis of representation collapse across downstream adaptation strategies.
    (a–c) Relative performance improvements of fine-tuning (FT), linear probing (LP), and multi-task fine-tuning (MFT) compared with training from scratch (Scratch) baseline.
    (d–f) and (g–i) t-SNE visualizations of atomic (node) and edge representations, respectively, projected into two-dimensional spaces.
    The clarity and separation of atomic and edge representation clusters are quantified using the Davies–Bouldin (DB) index, normalized with respect to the pretrained model.
    Larger relative DB index values indicate more severe degradation in cluster cohesion and separability, providing quantitative evidence of collapse in the pretrained representation space.
    }
    \label{fig:results}
  \end{center}
  \vskip -0.3in
\end{figure*}

In this work, we demonstrate that standard fine-tuning (FT) of pretrained atomistic models (Fig.~\ref{fig:overview}(b)) leads to a pronounced degradation in OOD generalization: 
the OOD performance of the fine-tuned model is substantially worse than its ID performance (Fig.~\ref{fig:results}(a)). 
We further show that this degradation is accompanied by a loss of the chemical prior knowledge learned during pretraining, indicative of representation collapse (Fig.~\ref{fig:results}(e)(h)). 
Although linear probing (LP) model (Fig.~\ref{fig:overview}(c)) yields a slightly better OOD performance than FT, it suffers a significant decrease in ID generalizability (Fig.~\ref{fig:results}(b)).
To address this issue, we propose multi-task fine-tuning (MFT), which jointly optimizes the downstream property-prediction task and the upstream force-field pretraining task (Fig.~\ref{fig:overview}(d)).
We systematically evaluate MFT using a  large atomistic model, DPA-3.1-3M~\cite{zhang2025graphneuralnetworkera}, pretrained on a diverse collection of force-field datasets spanning both materials and molecular systems~\cite{openlam-data-v1-web}. 
Under MFT, the relative improvement in OOD generalization (measured against training from scratch) matches the corresponding ID improvement and is comparable to the ID performance achieved by FT (Fig.~\ref{fig:results}(c)).
This improvement is accompanied by a clear mitigation of representation collapse (Fig.~\ref{fig:results}(f)(i)).
In terms of absolute accuracy, the MFT DPA-3.1-3M model achieves the best OOD performance across both molecular (\cref{table:molecule}) and materials (\cref{table:material}) benchmarks, including those designed around property-prediction-oriented architectures, demonstrating the robustness and practical readiness of the proposed approach for real-world molecular and materials design applications.

\section{Results}
\label{sec:results}

\subsection{Overview of MFT}

We provide an overview of different downstream adaptation strategies used in this work in Fig.~\ref{fig:overview}.
Each downstream adaptation setup is compared against training from scratch (Scratch) baseline, in which both the backbone and the task head are randomly initialized, as shown in Fig.~\ref{fig:overview}(a).
During FT, the pretrained backbone is transferred and coupled with a randomly initialized, task-specific prediction head, with all model parameters updated end-to-end, as depicted in Fig.~\ref{fig:overview}(b) .
Furthermore, we investigate the performance of LP model, i.e.~freezing the backbone of the pretrained model and fine-tuning only the task-specific head, as illustrated in Fig.~\ref{fig:overview}(c).
This setting constrains the representation to remain fixed and therefore free of distributional drift.

In this work, we propose MFT, as graphically displayed in Fig.~\ref{fig:overview}(d).
Similar to FT, the backbone of the pretrained model is transferred to the downstream task.
However, under MFT the backbone is connected to two task heads: one randomly initialized for downstream property prediction and the other inherited from the force-field prediction head used during pretraining.
Both task heads, together with the shared backbone, are jointly optimized on the downstream property objective and an auxiliary force-field objective.
The auxiliary pretraining task can be selected to align closely with the target downstream domain.
For example, when fine-tuning for molecular property prediction, the SPICE2 dataset~\cite{eastman2024nutmeg}, which is designed for training machine-learning interatomic potentials for organic molecules, can be used as the auxiliary force-field task.
For materials property prediction, the OMat24 dataset~\cite{barroso2024open}, which is proposed for universal materials machine-learning potentials with a particular emphasis on structural and compositional diversity, can be chosen as the corresponding force-field task.
This joint optimization strategy constrains the backbone to learn a representation space that is simultaneously optimal for the specific downstream property prediction task while preserving the broad chemical and configurational prior knowledge acquired during pretraining.
\cref{tab:method_compare} summarizes the comparison between these methods, while \cref{sec:method} provides a comprehensive description of each approach.

We investigate the fine-tuning behavior of the pretrained GNN model DPA-3.1-3M~\cite{zhang2025graphneuralnetworkera} on QM9 molecular property prediction tasks.
The results summarized in Fig.~\ref{fig:results} show that the performance gap between ID and OOD generalization is closely associated with representation collapse in the learned representation space.
A similar representation-collapse phenomenon is also observed when fine-tuning the pretrained JMP model~\cite{shoghi2023molecules}, with full results reported in Supplementary \cref{sec:jmpl}.
These observations indicate that representation collapse is a broadly occurring failure mode in fine-tuning pretrained atomistic models, rather than an artifact of a specific model architecture or pretraining setup.

\subsection{MFT closes the generalization gap}
\label{sec:results}

\textit{Data split.}
The QM9 dataset is partitioned into training, validation, ID test, and OOD test subsets.
The training, validation, and ID test sets are drawn from the same distribution, whereas OOD samples are selected from the tails of the property distributions (the lowest 10\% probability mass estimated via kernel density estimation), representing extreme or rare molecular properties.
This splitting protocol follows established OOD benchmarks for molecular property prediction, namely BOOM~\cite{antoniuk2025boom}.

\textit{Significant ID gains and OOD degradation under FT.}
Relative to training from scratch, FT yields an average improvement of 17.8\% in ID accuracy, but only a 5.6\% improvement in OOD accuracy, as reported in Fig.~\ref{fig:results}(a).
Across the eight property prediction tasks, the relative OOD improvement is consistently smaller than the corresponding ID improvement, with the sole exception of electronic spatial extent ($\langle R^2 \rangle$).
Notably, for the dipole moment ($\mu$) and the lowest unoccupied molecular orbital (LUMO) energy, the fine-tuned model underperforms the from-scratch baseline in the OOD regime by 8.7\% and 3.8\%, respectively.

\textit{OOD enhancement and ID degradation of LP.}
As reported in Fig.~\ref{fig:results}(b), the OOD accuracy of LP improves by 7.8\% relative to a model trained from scratch, which is marginally higher than the improvement achieved by FT.
In contrast, the ID generalizability decreases by 37\% compared to the from-scratch baseline, indicating that the expressive capacity of the model is severely limited by freezing the backbone.
This behavior is expected, as the target quantities predicted in downstream property prediction tasks differ substantially from those in force-field learning.
Specifically, property prediction models aim to infer quantities such as electronic structure properties (HOMO, LUMO, GAP, and dipole moment) and thermodynamic properties (e.g.~heat capacity), whereas force-field models are trained to predict ground-state energies and their derivatives with respect to atomic coordinates (forces).
This fundamental shift in prediction paradigm necessitates adaptation of the representation space.

\textit{MFT closes the generalization gap.}
An effective fine-tuning strategy must allow representations to adapt to downstream task requirements while simultaneously preserving the chemical and configurational prior knowledge acquired during pretraining, and these requirements are fulfilled by the proposed MFT strategy.
As presented in Fig.~\ref{fig:results}(c), under MFT the relative improvement in OOD generalization (measured against training from scratch) reaches 17.1\%, which closely matches the corresponding ID improvement (17.3\%) and is comparable to the ID performance achieved by FT (17.8\%).
In the best-performing case ($\langle R^2\rangle$), OOD generalization improves by 36.1\%, whereas in the worst case ($\mu$), OOD performance remains comparable to that of a model trained from scratch.
In this sense, MFT exhibits no discernible generalization gap, and its OOD generalizability approaches the theoretical upper bound given by the ID generalizability.

\begin{table}[t]
  \caption{
  Comparison of downstream adaptation strategies.
  The table highlights the differences between training from scratch (Scratch), fine-tuning (FT), linear probing (LP), and our proposed multi-task fine-tuning (MFT) in terms of backbone initialization, update parameters and the inclusion of auxiliary training tasks.
  }
  \label{tab:method_compare}
  \centering
  \scriptsize
  \begin{spacing}{1.2}
  \begin{tabular}{lcccc}
    \toprule
    \textbf{Method} & \textbf{\makecell{Backbone\\Initialization}} & \textbf{\makecell{Update\\Backbone}} & \textbf{\makecell{Update\\Task Head}} & \textbf{\makecell{Auxiliary\\Task}} \\
    \midrule
    Scratch & Random & $\surd$ & $\surd$ & $\times$ \\
    FT & Pretrained & $\surd$ & $\surd$ & $\times$ \\
    LP & Pretrained & $\times$ & $\surd$ & $\times$ \\
    \textbf{MFT (Ours)} & Pretrained & $\surd$ & $\surd$ & $\surd$ \\
    \bottomrule
  \end{tabular}
  \end{spacing}
\end{table}

\subsection{MFT mitigates the collapse of representations}

\textit{Pretrained atomic representations.} 
Fig.~\ref{fig:results}(d) visualizes the atomic (node) representations of the pretrained model.
Atoms of different chemical species are indicated by distinct colors, while atoms of the same species are further differentiated by their number of covalent neighbors (see Supplementary Section~\ref{sec:rep_detail} for details).
As shown in Fig.~\ref{fig:results}(d), the atomic representations of the pretrained model are well aligned with chemical intuition: atoms of the same species and identical covalent environments form well-separated clusters, clearly distinguished from atoms of different species or coordination environments.

\textit{Collapse of atomic representations.} 
After FT, however, substantial overlap emerges between representations of distinct species, as illustrated in Fig.~\ref{fig:results}(e).
For instance, the representations of nitrogen and oxygen atoms begin to intermingle.
In addition, the resolution with respect to local covalent environments is markedly reduced.
Carbon atoms with different numbers of covalent neighbors become mixed, oxygen atoms with two neighbors split into multiple clusters influenced by hydrogen and fluorine, and hydrogen atoms separate into clusters by single-bonded carbon cluster.
The clarity and separation of the representation clusters can be quantitatively assessed using the Davies–Bouldin (DB) index~\cite{dbindex1979}, where lower values indicate more distinct and cohesive clusters.
As shown in Fig.~\ref{fig:results}(e), after FT, the DB index of the atomic representations increases by a factor of 6.2 relative to the pretrained representations, indicating a substantial blurring and loss of cluster separability.
Together, these observations indicate a collapse of the chemically meaningful prior representations, both in chemical identity and local covalent environment, learned during pretraining after FT.

\textit{MFT mitigates the collapse of atomic representations.} 
The OOD improvement of MFT is accompanied by a clear mitigation of the collapse of atomic representation.
As illustrated in Fig.~\ref{fig:results}(f), atomic representations corresponding to the same chemical species and covalent environments are clearly separated into distinct clusters.
Only minor exceptions are observed: nitrogen atoms with three covalent neighbors are split into two clusters influenced by hydrogen.
The DB index, which quantitatively measures cluster separation, increases by only 30\%, in stark contrast to standard FT, where the DB index increases by 520\% relative to that of the pretrained atomic representations.


\textit{Pretrained edge representations.}
As a GNN, DPA-3.1-3M employs edge representations to encode pairwise atomic relationships within a preset cutoff radius.
Fig.~\ref{fig:results}(g) visualizes the edge representations of carbon–hydrogen (C–H) and carbon–oxygen (C–O) pairs, two prevalent edge types in the downstream tasks.
These edges are further stratified into \emph{near}, \emph{middle}, and \emph{far} categories according to their pairwise Euclidean distances: \emph{near} corresponds to atom pairs forming covalent bonds, while \emph{middle} and \emph{far} denote nonbonded pairs whose features are updated with or without angular information, respectively.
The pretrained model exhibits a highly organized edge representation, with edges distinctly clustered by both chemical identity and interatomic distance, indicating that the pretrained model encodes strong chemical and geometric priors.

\textit{Collapse of  edge representations.}
In contrast, FT induces an evident structural degradation of this latent organization, as shown in Fig.~\ref{fig:results}(h).
Although \emph{near} edges remain relatively cohesive due to strong covalent signals, the representations of \emph{middle} and \emph{far} edges increasingly overlap: C–O edge representations begin to mix with C–H representations, and the \emph{middle} and \emph{far} subclasses become entangled.
This loss of configurational resolution is quantitatively reflected by a 1.3-fold increase in the DB index.
Overall, the observed collapse indicates that the model’s ability to distinguish long-range spatial correlations is compromised, suggesting that FT prioritizes downstream label fitting at the expense of preserving chemically and geometrically meaningful representations.

\textit{MFT mitigates the collapse of edge representations.}
The proposed MFT method effectively mitigates the collapse of edge representation.
As shown in Fig.~\ref{fig:results}(i), edge representations corresponding to different pair types and interatomic distances remain well separated under MFT, with a DB index that is even 10\% lower than that of the pretrained edge representations. This behavior contrasts sharply with FT, which leads to a 30\% increase in the DB index.

\begin{table}[t]
  \caption{
  Quantification of distribution shift between pretrained and fine-tuned representations using 1-Wasserstein distance ($\mathcal{W}_{\textrm{Pretrain}\rightarrow\textrm{FT}}$, $\mathcal{W}_{\textrm{Pretrain}\rightarrow\textrm{MFT}}$). 
  Metrics are reported for the QM9 LUMO prediction task across training, ID, and OOD sets.
  }
  \label{tab:w1}
  \begin{center}
        \begin{tabular}{lccc}
          \toprule
            & Train & ID & OOD\\
          \midrule
          $\mathcal{W}_{\textrm{Pretrain}\rightarrow\textrm{FT}}$  & 3.76 & 3.75 & 6.54 \\
         $\mathcal{W}_{\textrm{Pretrain}\rightarrow\textrm{MFT}}$  &  1.51 & 1.49 & 1.97 \\
          \bottomrule
        \end{tabular}
  \end{center}
  \vskip -0.1in
\end{table}

\subsection{MFT mitigates distribution shift between pretrained and fine-tuned representations}

\textit{Significant representation drift under FT.}
We formally quantify the similarity, or equivalently the drift, between representation distributions of the pretrained and fine-tuned models using the 1-Wasserstein distance, denoted by $\mathcal W$.
As shown in Table~\ref{tab:w1}, the drift distances for the training and ID test distributions before and after FT are very close, which is expected since they are sampled from the same underlying distribution and therefore undergo similar modifications.
In contrast, a significantly larger drift is observed for the OOD test distribution, indicating that FT induces substantially stronger representation changes for OOD samples than for ID samples.
This pronounced distributional shift provides clear evidence of overfitting induced by FT.

\textit{MFT mitigates representation drift.}
MFT significantly reduces the distribution shift between pretrained and fine-tuned representations.
As reported in \cref{tab:w1}, under the MFT framework the representation distribution drifts for the training and ID test sets are 1.51 and 1.49, respectively, which are substantially smaller than those observed under FT (3.76 and 3.75, respectively).
The OOD distribution drift is 1.97, which is only mildly larger than the corresponding training/ID drifts and remains significantly smaller than the OOD distribution drift (6.54) induced by FT.

However, eliminating representation drift entirely is counterproductive. 
As shown in \cref{sec:results}, LP which maintains zero representation drift by freezing the backbone, suffers from poor ID performance. 
This occurs because the model fails to bridge the paradigm shift between force-field pretraining and downstream property prediction. 
An optimal fine-tuning strategy must strike a balance: adapting representations to specific task requirements while preserving the foundational chemical priors learned during pretraining. 
Our results suggest that MFT successfully achieves this equilibrium.

\begin{table*}[t]
  \notsotiny
  \caption{
  Root-mean-square error (RMSE) on eight molecular property prediction tasks from the QM9 BOOM benchmark.
    We compare our methods, DPA3-Scratch (training from scratch), DPA3-LP (linear probing), DPA3-FT (standard fine-tuning), and DPA3-MFT (multi-task fine-tuning), against ChemBERTa~\cite{chithrananda2020chemberta}, Graphormer~\cite{ying2021graphormer}, EGNN~\cite{victor2021egnn}, ET~\cite{tholke2021equivariant}, MoLFormer~\cite{Ross2022molformer}, RT~\cite{born2023regression}, Geoformer~\cite{wang2023geoformer}, Chemprop~\cite{heid2023chemprop}, MACE~\cite{mace2023jcp}, ModernBERT~\cite{modernbert}, and GotenNet~\cite{aykent2025gotennet}.
    The best and second-best results are highlighted in \textbf{bold} and \underline{underlined}, respectively.
    Black text denotes the best in-distribution (ID) performance, while \textcolor{blue}{blue} text denotes the best out-of-distribution (OOD) performance.
    All DPA3 results are averaged over three independent training runs.
  }
  \label{table:molecule}
  \begin{center}
  \begin{spacing}{1.2}
  \begin{tabular}{L{1.2cm} | C{1.0cm} | C{0.5cm} | C{1.3cm} C{1.3cm} C{1.3cm} C{1.5cm} C{1.0cm} C{1.1cm} C{1.1cm} C{1.1cm}}
    \toprule
        Model &\#parameters & Split & HOMO & LUMO & GAP & ZPVE & $\langle R^2 \rangle$ & $\alpha$ & $\mu$ & $C_v$ \\ \hline

        \multirow{2}{*}{ChemBERTa} & \multirow{2}{*}{83M} & ID & 0.0070 & 0.0093 & 0.0104& 0.00390 & 51.7 & 1.254 & 0.713 & 0.489\\
        && OOD & 0.0244 & 0.0256 & 0.0309 & 0.02253 & 302.6 & 6.328 & 2.723 & 2.765 \\ \hline

        \multirow{2}{*}{Graphormer} 
        & \multirow{2}{*}{47.1M} & ID  & 0.0040 & 0.0042 & 0.0055 & 0.00024 & 33.4 & 0.431 & 0.626 & 0.180\\
        && OOD & 0.0236 & 0.0256 & 0.0275 & 0.01990 & 314.1 & 7.356 & 2.232 &3.667\\ \hline

        \multirow{2}{*}{EGNN} & \multirow{2}{*}{0.2M} & ID  & 0.0048 & 0.0052 & 0.0069& 0.00103 &19.6 &0.566& 0.481 &0.267 \\
        && OOD  & 0.0212& 0.0236& 0.0312 &0.00583 &181.3 &5.659 &2.446 &2.079\\ \hline

        \multirow{2}{*}{ET} & \multirow{2}{*}{6.8M} & ID & 0.0027 & 0.0031 & 0.0043 & 0.00057 & 28.2 & 0.490 & 0.368 & 0.160\\
        && OOD & 0.0220 & 0.0236 & 0.0271 & 0.01710 & 298.0 & 6.568 & 2.257 & 3.405\\ \hline

        \multirow{2}{*}{MoLFormer} & \multirow{2}{*}{48M} & ID &0.0050 & 0.0052 & 0.0064 & 0.00106 & 40.1 & 1.047 & 0.602 & 0.785 \\
        && OOD & 0.0236 & 0.0256 &  0.0275 & 0.01990 & 314.1 & 7.356 & 2.232 & 3.667 \\ \hline

        \multirow{2}{*}{RT} & \multirow{2}{*}{27M} & ID &0.0090 & 0.0102 & 0.0133 & 0.00289 & 68.2 & 2.264 & 1.104 & 0.654\\
        && OOD & 539.74 & 584.88 & 0.0339 & 27.6906 & 25458 & 69968 & 2.719 & 11435\\ \hline

        \multirow{2}{*}{Geoformer} & \multirow{2}{*}{51M} & ID  & 0.0027 & 0.0028 & 0.0046 & 0.00030 & 10.9 & 0.326 & 0.847 & 0.124\\
        && OOD  & 0.0157 & 0.0186 & 0.0240 & 0.00557 & 63.8 & 4.201 & 2.544 &1.354\\ \hline

        \multirow{2}{*}{Chemprop} & \multirow{2}{*}{0.2M} & ID & 0.0041 &0.0046 
        & 0.0058 & 0.00188 & 35.8 & 0.866 & 0.545 & 0.340\\
        && OOD & 0.0189 &0.0179 &0.0269 &0.01277 &233.7 &4.850 &2.304 &2.118\\ \hline 

        \multirow{2}{*}{MACE} & \multirow{2}{*}{3.9M} & ID  & 0.0150 & 0.0135 & 0.0181 &0.00180 &9.8& 0.322 &0.430 &0.134\\
        && OOD  &0.0339 &0.0247 &0.0409& 0.00210 &68.3 & \ul{1.543} & 2.228 & 0.229\\ \hline

        \multirow{2}{*}{ModernBERT} & \multirow{2}{*}{111M} & ID & 0.0064 & 0.0076& 0.0095 & 0.00073 & 40.1 & 0.870 & 0.698 & 0.407 \\
        && OOD  & 0.0216 & 0.0232 & 0.0324 & 0.00170 & 228.7 & 2.489 & 2.657 &0.611 \\ \hline

        \multirow{2}{*}{GotenNet} & \multirow{2}{*}{9.2M} & ID  & 0.0052 & 0.0039 & 0.0088 & 0.00043 & 11.8 & 0.553 & 0.319 & 0.197\\
        && OOD & \ul{0.0126} & \ul{0.0118} & 0.0229 & 0.00053 & 16.7 & 1.825 & 2.173 & 0.302\\  \hline \hline


        \multirow{2}{*}{DPA3-Scratch} &\multirow{2}{*}{3.3M}& ID & 0.0030~(0.0000) & 0.0029~(0.0000) & 0.0045~(0.0000) &  0.00013~(0.00000) & 8.5~(0.3) & 0.365~(0.006) & 0.294~(0.003) & 0.116~(0.001) \\
        && OOD & 0.0155~(0.0004) & 0.0130~(0.0002) & 0.0244~(0.0003) & 0.00029~(0.00000) & 21.4~(0.8) & 1.907~(0.019) & \ul{2.087}~(0.003) & 0.200~(0.006) \\ \hline

        \multirow{2}{*}{DPA3-LP} &\multirow{2}{*}{3.3M}& ID & 0.0046~(0.0000) & 0.0054~(0.0000) & 0.0077~(0.0000) & 0.00016~(0.00000) & 8.8~(0.0) & 0.458~(0.000) & 0.309~(0.000) & 0.149~(0.000) \\
        && OOD & \textcolor{blue}{\textbf{0.0125}}~(0.0001) & 0.0128~(0.0002) & \ul{0.0193}~(0.0001) & 0.00029~(0.00000) & \ul{14.2}~(0.0) & 1.641~(0.010) & \textcolor{blue}{\textbf{1.747}}~(0.011) & 0.281~(0.002) \\ \hline

        \multirow{2}{*}{DPA3-FT} &\multirow{2}{*}{3.3M}& ID & \underline{0.0022}~(0.0000) & \textbf{0.0023}~(0.0000) & \textbf{0.0034}~(0.0000) &  \textbf{0.00010}~(0.00000) & \textbf{8.1}~(0.0) & \underline{0.297}~(0.001) & \textbf{0.276}~(0.001) & \textbf{0.097}~(0.001) \\
        && OOD & 0.0156~(0.0001) & 0.0135~(0.0003) & 0.0201~(0.0002) & \ul{0.00026}~(0.00000) & 18.7~(2.1) & 1.639~(0.022) & 2.267~(0.018) & \ul{0.188}~(0.002) \\ \hline

        \multirow{2}{*}{DPA3-MFT} &\multirow{2}{*}{3.3M}& ID & \textbf{0.0021}~(0.0000) & \underline{0.0024}~(0.0000) & \textbf{0.0034}~(0.0000) &  \textbf{0.00010}~(0.00000) & \underline{8.3}~(0.0) & \textbf{0.296}~(0.001) & \underline{0.278}~(0.001) & \textbf{0.097}~(0.001) \\
        && OOD & 0.0134~(0.0002) & \textcolor{blue}{\textbf{0.0107}}~(0.0002) & \textcolor{blue}{\textbf{0.0171}}~(0.0002) & \textcolor{blue}{\textbf{0.00025}}~(0.00000) & \textcolor{blue}{\textbf{13.7}}~(0.1) & \textcolor{blue}{\textbf{1.536}}~(0.005) & \ul{2.087}~(0.008) & \textcolor{blue}{\textbf{0.181}}~(0.001) \\

    \bottomrule
  \end{tabular}
  \end{spacing}
  \end{center}
  \vskip -0.1in
\end{table*}

\subsection{MFT outperforms SOTA molecular property predictors}

The proposed MFT strategy is benchmarked against state-of-the-art (SOTA) property prediction baselines on the BOOM QM9 benchmark, as summarized in \cref{table:molecule}.
The MFT model (denoted as DPA3-MFT) is compared with standard fine-tuning (DPA3-FT) and linear probing (DPA3-LP) applied to the same pretrained backbone, namely DPA-3.1-3M; a schematic illustration of these settings is provided in Fig.~\ref{fig:overview}.
For a fair comparison, the same model architecture is also trained from scratch (DPA3-Scratch).
Both DPA3-FT and DPA3-MFT demonstrate a clear advantage in ID scenarios: they consistently achieve the best ID performance across all eight tasks, confirming that fine-tuning from a pretrained model outperforms SOTA task-specific training from scratch, and that the MFT strategy does not compromise ID accuracy.
The OOD results, evaluated using property extrapolation splits following BOOM~\cite{antoniuk2025boom}, highlight the primary strength of the MFT approach.
DPA3-MFT achieves the best OOD performance in six out of the eight tasks and the second-best performance in one task.
In contrast, DPA3-FT often exhibits a performance trade-off when generalizing to OOD data, likely due to the collapse of chemical priors encoded in the representations learned during pretraining, resulting in elevated OOD errors.
By explicitly mitigating this effect, DPA3-MFT substantially improves robustness under distribution shift.

\begin{table*}[t]
  \notsotiny
  \caption{Mean absolute error (MAE) of different GNN models on the dielectric, elasticity, and perovskite datasets from MatBench.
    Evaluation includes 50-fold cross-validation over five distinct out-of-distribution (OOD) settings introduced in Ref.~\cite{sadman2024oodmatbench} (with detailed descriptions of the OOD splitting strategies provided in \cref{sec:matbench_detail}), as well as a 5-fold random split for in-distribution (ID) comparison.
    We benchmark DPA3-FT (standard fine-tuning) and DPA3-MFT (multi-task fine-tuning) against CGCNN~\cite{prl2018cgcnn}, MEGNet~\cite{chi2019megnet}, SchNet~\cite{schutt2017schnet, schutt2017quantum}, DimeNet++~\cite{gasteiger_dimenetpp_2020}, ALIGNN~\cite{choudhary2021atomistic}, DeeperGATGNN~\cite{OMEE2022100491}, coGN~\cite{ruff2024connectivity}, and coNGN~\cite{ruff2024connectivity}.
    Best and second-best performances are highlighted in \textbf{bold} and \underline{underlined}, respectively.
  }
  \label{table:material}
  \begin{center}
  \begin{spacing}{1.2}
  \begin{tabular}{c | c| c  c  c   c   c  c  c  c | c c }
    \toprule
    OOD Split & Datasets & CGCNN & MEGNet & SchNet & DimeNet++ & ALIGNN & DeeperGATGNN & coGN & coNGN & DPA3-FT & DPA3-MFT \\ \hline

    \multirow{3}{*}{LOCO} & dielectric & 0.5144 & 2.6830 & 2.7074 & 2.7720 & 0.8592 & 0.5911 & 0.4984 & \underline{0.4983} & 0.5305 & \textbf{0.4690}\\

    & elasticity  & \textbf{0.0585} & 1.4468 & 1.4065 & 1.4242 & 0.0974 & 0.1173 & 0.1017 &0.1019 & 0.0806 & \underline{0.0774} \\

    & perovskites & 0.0651 & 1.4654 & 1.4644 & 1.4666 & 0.0386 & 
    0.0365 & 0.0631 & 0.0632 & \underline{0.0350} & \textbf{0.0341}\\ \hline

    \multirow{3}{*}{SparseXcluster} & dielectric & 0.6006 & 2.8044 & 2.8368 & 2.9378 & 0.6953 & 1.4056 & \underline{0.5241} & 0.5242 & 0.6069 & \textbf{0.4806} \\

    &elasticity & \textbf{0.0499} & 1.4113 & 1.3455 & 1.3562 & 0.0834 & 0.1109 & 0.1416 & 0.1417 & 0.0754 & \underline{0.0689}\\

    &perovskites& 0.0839 & 1.4485 & 1.4509 & 1.4567 & 0.0468 & 0.0464 & 0.1306 & 0.1306 & \textbf{0.0384} & \underline{0.0403}\\ \hline

    \multirow{3}{*}{SparseXsingle} & dielectric & 0.9888 & 3.8384 & 3.9767 & 3.8629 & 1.5115 & 1.5755 & \underline{0.5519} & \textbf{0.5518} & 1.5335 & 1.2463\\

    & elasticity & 0.0895 & 1.3099 & 1.2363 & 1.3214 & 0.0853 & 0.1140 & 0.2919 & 0.2918 & \underline{0.0783} & \textbf{0.0782}\\

    & perovskites & 0.0689 & 1.5005 & 1.5133 & 1.5248 & 0.0457 & 0.0373 & 0.2199 & 0.2200 & \underline{0.0370} & \textbf{0.0358}\\ \hline

    \multirow{3}{*}{SparseYcluster} & dielectric & 0.5254 & 2.5918 & 2.5988 & 2.5947 & 0.4359 &  0.3959 & 0.4673 & 0.4674 & \underline{0.3354} & \textbf{0.2688}\\

    & elasticity & 0.0752 & 1.5659 & 1.5592 & 1.5454 & 0.0631 & 0.0858 & 0.0823 & 0.0823 & \underline{0.0529} & \textbf{0.0527} \\

    & perovskites & 0.0740 & 1.4685 & 1.4736 & 1.4732 & 0.0341 & 0.0333 & 0.0925 & 0.0926 & \underline{0.0322} & \textbf{0.0300}\\ \hline

    \multirow{3}{*}{SparseYsingle} & dielectric & 0.4777 & 2.5706 & 2.5566 & 2.5866 & 0.2513 & 0.2733 & 0.3589 & 0.3589 & \underline{0.2008} & \textbf{0.1790}\\

    & elasticity & 0.0840 & 1.4491 & 1.4855 & 1.4828 & \underline{0.0450} & 0.0807 & 0.1852 & 0.1853 & 0.0477 &\textbf{0.0435}\\
    
    & perovskites & 0.0768 & 1.5048 & 1.5173 & 1.5148 & 0.0243 & 0.0259 & 0.1884 & 0.1886 & \underline{0.0235} &\textbf{0.0212}\\  \hline

    \multirow{3}{*}{Random} & dielectric & 0.5988 & 0.3391 & 0.3277 & 0.3400 & 0.3449 & 0.3355 & 0.3088 & 0.3142 & \underline{0.3082} & \textbf{0.2639}\\
    & elasticity & 0.0895 & 0.0871 & 0.0796 & 0.0792 & 0.0715 & 0.0903 & 0.0689 & 0.0670 & \underline{0.0627} & \textbf{0.0609}\\
    & perovskites & 0.0452 & 0.0352 & 0.0342 & 0.0376 & 0.0288 & 0.0288 & \textbf{0.0269} & 0.0290 & 0.0272 &\underline{0.0270}\\ 

    \bottomrule
  \end{tabular}
  \end{spacing}
  \end{center}
  \vskip -0.1in
\end{table*}

\subsection{MFT outperforms SOTA material property predictors}

To evaluate the robustness of the proposed MFT strategy on inorganic materials property prediction tasks, we conducted a comprehensive benchmark comprising an ID random split and five OOD splits constructed using structure- and property-based criteria~\cite{sadman2024oodmatbench}.
These splitting strategies were applied to three datasets from MatBench~\cite{dunn2020benchmarking}: refractive index (dielectric), shear modulus (elasticity), and perovskite formation energy (perovskites).
A summary of the three datasets is provided in \cref{tab:matbench_info}.
The OOD split strategies leverage the Orbital-Field Matrix (OFM)~\cite{pham2017machine} representation to encode electronic structure and orbital interactions, yielding a physically interpretable feature space for subsequent clustering and density estimation~\cite{karamad2020orbital}.
Specifically, the five OOD splitting methods include Leave-One-Cluster-Out (LOCO), which partitions the dataset into 50 folds via $k$-means clustering in OFM feature space.
In addition, four sparse-target strategies (SparseXsingle, SparseXcluster, SparseYsingle, and SparseYcluster) identify samples with the lowest density either in feature space (X) or property space (Y) using Gaussian kernel density estimation.
These low-density samples are further grouped into 50 clusters using $k$-means clustering, and OOD test sets are constructed by selecting either a single representative point from each cluster (single) or the representative point together with its 10 nearest neighbors (cluster), resulting in 50 OOD folds.
For each splitting method, the 50 OOD test sets are generated in a leave-one-out manner, where one fold is held out as the OOD test set while the remaining data are used for training and validation.
Further details of the five OOD split strategies are provided in Supplementary \cref{sec:matbench_detail}.

As demonstrated in \cref{table:material}, DPA3-MFT achieves the best performance in 13 of the 18 evaluated tasks and ranks second in 4 additional tasks.
In nearly all cases, DPA3-MFT outperforms the standard fine-tuning baseline (DPA3-FT), with an average improvement of 7.8\%; the sole exception occurs for the perovskite SparseXcluster split.
Compared with other SOTA models, DPA3-MFT maintains a consistent performance advantage, although certain competing methods benefit from task-specific architectural design.
Notably, DPA3-MFT leads OOD performance across all property-outlier-based split strategies (SparseYsingle and SparseYcluster).
Exceptions arise for structure-based OOD splits, including elasticity under LOCO and SparseXcluster, and dielectric under SparseXsingle.
In particular, the relatively poor performance of DPA3-MFT on the dielectric SparseXsingle split is attributable to an exceptionally high MAE in one of the 50 folds as shown in Fig.~\ref{fig:dielectric}, which disproportionately inflates the average test error.
The most pronounced gains from MFT are observed on the dielectric dataset, where DPA3-MFT improves upon DPA3-FT by 16.0\% (compared with 4.1\% and 3.1\% improvements for elasticity and perovskites, respectively).
This behavior likely arises because the perovskites and elasticity tasks predict perovskite formation energies and shear moduli (the latter being closely related to second derivatives of the energy), which are intrinsically aligned with the energy and force prediction objectives used during pretraining.
In contrast, the dielectric (refractive index) task probes electronic structure properties, representing a substantially larger task-domain shift relative to force-field pretraining.
As a result, successful adaptation requires more substantial representation updates, making the preservation of chemically meaningful priors acquired during pretraining particularly critical; MFT is therefore especially effective in this regime.

\begin{figure*}[t]
  \begin{center}
    \centerline{\includegraphics[width=2\columnwidth]{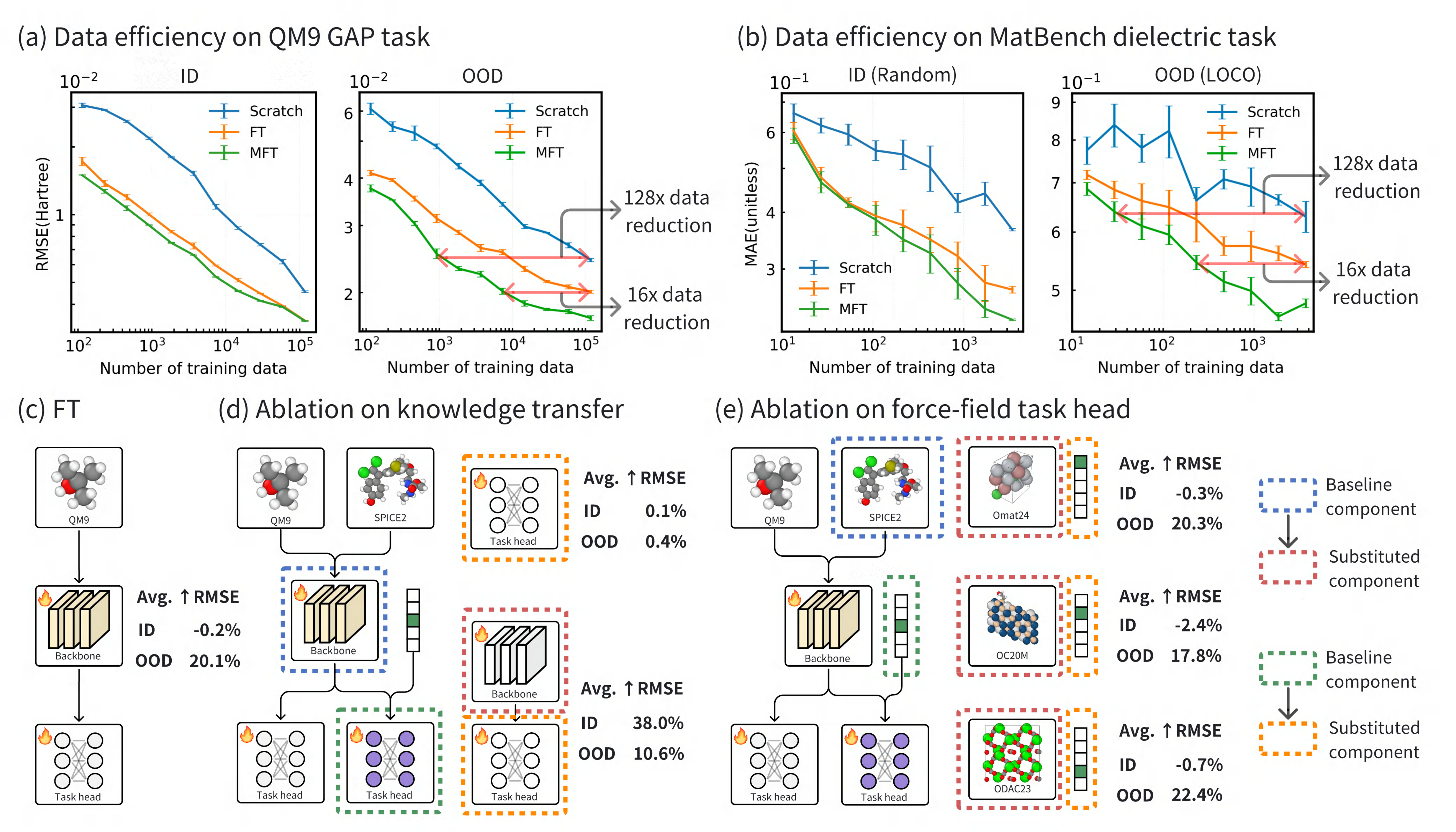}}
    \caption{
Data efficiency and ablation study of multi-task fine-tuning (MFT).
(a) Data efficiency on the QM9 band-gap (GAP) prediction task.
(b) Data efficiency on the MatBench dielectric prediction task.
In (a,b), MFT is compared with training from scratch (Scratch) and standard fine-tuning (FT); in-distribution (ID) and out-of-distribution (OOD) test RMSEs or MAEs are shown as functions of training set size.
(c–e) Ablation study of MFT on the QM9 HOMO, LUMO, and GAP prediction tasks.
(c) FT baseline.
(d) Ablation of transferred pretrained components.
(e) Ablation of auxiliary force-field task selection.
``Avg. $\uparrow$RMSE'' denotes the average RMSE increase across the HOMO, LUMO, and GAP prediction tasks relative to the baseline MFT configuration.}
    \label{fig:data_effi_and_ablation}
  \end{center}
  \vskip -0.3in
\end{figure*}

\subsection{MFT enhances data efficiency and few-shot OOD generalization}

To evaluate the impact of downstream data size on model performance, we investigated the data efficiency of MFT using the QM9 band gap (GAP) prediction task and MatBench refractive index (dielectric) prediction task.
As depicted in Fig.~\ref{fig:data_effi_and_ablation}(a-b), MFT is compared against two baselines: training from scratch and FT.
In these experiments, the size of the training set is varied by randomly down-sampling the original training data to as few as 114 data points for GAP and 14 data points for dielectric.
The ID and OOD test root-mean-square errors (RMSEs) or mean absolute errors (MAEs) are plotted as functions of the training data size, which spans three orders of magnitude.
The implementation details are provided in Supplementary \cref{sec: data_effi_detail}.

Fig.~\ref{fig:data_effi_and_ablation}(a) illustrates the results of QM9 GAP prediction task.
In terms of ID accuracy, both FT and MFT consistently outperform training from scratch across all data scales.
MFT maintains a modest but consistent advantage over FT, although this gap narrows when the training data size reaches 50\% of the full dataset.
In contrast, for OOD accuracy, MFT significantly outperforms both the from-scratch and FT baselines across all training sample sizes, demonstrating a strong few-shot generalization capability.
Notably, with only $1/128$ of the training data, MFT achieves an OOD accuracy comparable to that of a from-scratch model trained on the full dataset.
Moreover, MFT requires only $1/16$ of the training data to match the OOD performance of an FT model trained using the full dataset.

The MatBench dielectric prediction results, as illustrated in Fig.~\ref{fig:data_effi_and_ablation}(b), further validate these findings. 
For ID accuracy, both MFT and standard FT consistently outperform training from scratch across all data regimes. 
Although MFT maintains a performance advantage over FT, this gap narrows in the extremely low–data regime.
For OOD performance, MFT substantially outperforms both FT and training from scratch at all training set sizes.
Notably, MFT exhibits exceptional data efficiency, achieving OOD accuracy comparable to from-scratch and FT models trained on the full dataset using only $1/128$ and $1/16$ of the data, respectively.
These trends closely mirror those observed for the GAP task.

Given that labeled data for molecular and materials property prediction is typically sparse and computationally expensive to generate, for example, accurate band gaps often require costly hybrid exchange-correlation functionals, this robust performance across a wide range of data regimes highlights the practical utility of MFT for real-world molecular and materials design applications.

\subsection{Ablation Study}

We conducted ablation experiments on the QM9 HOMO, LUMO, and GAP prediction tasks to disentangle the contributions of knowledge transfer and to assess the impact of auxiliary task selection.
The baseline MFT configuration uses SPICE2 as the auxiliary force-field task and transfers both the pretrained backbone and the force-field task head.
We further consider variants in which the pretrained task head is not transferred, as well as variants in which neither the pretrained backbone nor the pretrained task head is transferred.
In addition, we examine the sensitivity of MFT performance to the choice of auxiliary force-field task by replacing SPICE2 with datasets from other research domains, including OMat24 (inorganic materials), OC20M (heterogeneous catalysis), and ODAC23 (metal–organic frameworks, MOFs).

As illustrated in Fig.~\ref{fig:data_effi_and_ablation}(d), transferring the force-field task head is not strictly necessary, resulting in only marginal increases in RMSE of 0.1\% (ID) and 0.4\% (OOD) when the task head is replaced with a randomly initialized one.
By contrast, omitting transfer of both the pretrained backbone and the task head markedly degrades performance, increasing RMSE by 38.0\% in ID and 10.6\% in OOD relative to the baseline.
Notably, even without explicit knowledge transfer, the OOD performance achieved by joint multi-task training remains superior to that of FT as shown in Fig.~\ref{fig:data_effi_and_ablation}(c), highlighting the intrinsic effectiveness of the multi-task framework in aligning downstream property prediction with the chemical and configurational priors encoded by force-field training tasks.

We further investigated the influence of domain alignment between the auxiliary force-field task and the downstream property prediction task.
A clear negative transfer effect is observed when the auxiliary domain diverges substantially from the target task.
The results are summarized in Fig.~\ref{fig:data_effi_and_ablation}(e).
While the ID performance remains relatively stable, the OOD performance degrades markedly across all three tasks, with average RMSE increases of 20.3\%, 17.8\%, and 22.4\% when using OMat24, OC20M, and ODAC23 as auxiliary tasks, respectively.
Notably, the OOD performance obtained with OMat24 and ODAC23 is even inferior to that of FT.
In contrast, OC20M yields a slight improvement in OOD performance relative to FT, which can likely be attributed to the partial structural overlap between catalytic adsorbates and the small organic molecules in QM9, suggesting that limited structural similarity can still benefit OOD generalization.
These observations are consistent with prior findings in the literature: knowledge transfer is not universally beneficial, and indiscriminate transfer across mismatched domains can induce \textit{negative transfer}~\cite{chen2019catastrophic}, wherein unrelated auxiliary data degrades downstream model performance.

\section{Discussion}

Pretrained atomistic models offer a promising unified foundation for molecular and materials property prediction; however, their practical impact critically depends on reliable generalization beyond the training domain.
In this work, we identify and systematically analyze a critical yet underexplored failure mode of fine-tuning pretrained atomistic models -- \textit{representation collapse}.
This phenomenon involves the degradation of chemically and geometrically meaningful priors acquired during pretraining, leading to a substantial deterioration in out-of-distribution (OOD) generalization performance.

To address this challenge, we propose \textit{multi-task fine-tuning} (MFT), which enables task-specific adaptation while preserving general-purpose chemical and configurational priors by jointly optimizing the downstream property-prediction objective together with an auxiliary force-field task inherited from pretraining.
Across a wide range of OOD settings, MFT consistently mitigates representation collapse and achieves state-of-the-art OOD performance without compromising in-distribution accuracy.
Moreover, the strong data efficiency of MFT highlights its practical relevance for real-world molecular and materials discovery, where labeled data are costly to obtain and extrapolation beyond known regimes is the norm rather than the exception.

Our analysis provides mechanistic insight into both the failure of standard fine-tuning and the success of MFT.
We show that OOD performance is strongly correlated with the degree of representation drift from the pretrained model, and that effective adaptation requires a careful balance between plasticity (needed to fit downstream targets) and stability (needed to retain pretrained knowledge).
Linear probing enforces stability but lacks expressiveness, whereas standard fine-tuning offers greater expressiveness at the cost of representation collapse.
By contrast, MFT occupies a favorable middle ground, enabling controlled representation adaptation guided by physically meaningful auxiliary objectives.

We anticipate that these insights will inform the design of next-generation atomistic models for molecular and materials discovery, where reliable extrapolation to previously unexplored chemical regimes is essential.
By enabling fine-tuning without forgetting, the proposed framework advances atomistic foundation models toward fulfilling their role as dependable engines for data-efficient and predictive scientific discovery.

\section{Methods}
\label{sec:method}

This section describes the pretraining procedure and the fine-tuning strategies, both standard fine-tuning and multi-task fine-tuning, employed in this work.

\subsection{Pretraining}

In this work, the large atomistic model DPA-3.1-3M~\cite{zhang2025graphneuralnetworkera} is pretrained on the OpenLAM-v1 collection~\cite{openlam-data-v1-web}, which consists of datasets labeled with energies and atomic forces computed using density functional theory (DFT)~\cite{hohenberg1964inhomogeneous, kohn1965self}. 
The OpenLAM-v1 collection spans a broad range of systems, including large-scale materials datasets such as OMat24~\cite{barroso2024open}, molecular datasets such as SPICE2~\cite{eastman2024nutmeg}, as well as other publicly available datasets contributed by the community. 
A complete list of datasets included in OpenLAM-v1 is provided in Ref.~\cite{peng2025lambench}.
We denote the datasets in the collection by $\{D_m\}_{m=1}^M$, where the OpenLAM-v1 collection comprises a total of $M=\text{31}$ datasets.
Each dataset $D_m = \{ (x_{mk}, E_{mk}, F_{mk}) \}_{k=1}^{K_m}$ contains $K_m$ configurations, where each data point consists of a three-dimensional atomic structure $x_{mk}$, its DFT-computed ground-state potential energy $E_{mk}$, and the corresponding atomic forces $F_{mk}$.
Formally, each structure is represented as $x_{mk} = (R_{mk}, Z_{mk})$, where $R_{mk} \in \mathbb{R}^{N_{mk} \times 3}$ denotes the Cartesian coordinates of the $N_{mk}$ atoms, and $Z_{mk} \in \mathbb{N}^{N_{mk}}$ denotes their atomic numbers.
For structures under periodic boundary conditions, the simulation cell tensor is included in the structural representation $x_{mk}$, and the corresponding DFT-computed virial tensor is included in the labels.
For simplicity of exposition, we do not explicitly discuss the cell tensor or the virial tensor in the remainder of this work.

The GNN backbone used to learn atomistic representations is denoted by $f_\theta$, with trainable parameters $\theta$, while the task-specific fitting network is denoted by $g_\phi$, with trainable parameters $\phi$.
The predicted potential energy is given by
\begin{equation}
    \hat{E}_{mk} = g_\phi \bigl(f_{\theta}(x_{mk}),\, c_m\bigr),
\end{equation}
where $c_m$ denotes the one-hot encoding of the dataset $D_m$.
Atomic forces are obtained as the negative gradient of the predicted potential energy with respect to the atomic coordinates,
\begin{equation}
    \hat{F}_{mk} = -\nabla_{R_{mk}} \hat{E}_{mk}.
\end{equation}
The empirical pretraining objective is defined as
\begin{align}\label{eq:pretrain-loss}\nonumber
&\mathcal L_{\mathrm{pre}}(\theta, \phi) \\
&= \sum_{m\in \mathcal{B}_{c}} \sum_{k\in \mathcal{B}_m}
\lambda_E \mathcal{L}_E(\hat{E}_{mk}, E_{mk})
+ \lambda_F \mathcal{L}_F(\hat{F}_{mk}, F_{mk}),
\end{align}
where $\mathcal{B}_{c}$ denotes a mini-batch of datasets randomly sampled from the dataset collection, and $\mathcal{B}_m$ denotes a mini-batch of configurations randomly sampled from dataset $D_m$.
Here, $\mathcal{L}_E$ and $\mathcal{L}_F$ are error metrics that quantify the discrepancy between the predicted and DFT reference \textit{energies} and 
\textit{forces}, respectively, while $\lambda_E$ and $\lambda_F$ are weighting coefficients that balance the contributions of the two loss terms.
The pretraining task is thus defined as minimizing the loss in Eq.~\eqref{eq:pretrain-loss}:
\begin{flalign}
\mathrm{(Pretraining)}&&
\theta_0, \phi_0 =
\underset{\theta, \phi}{\arg\min}
\; \mathcal{L}_{\mathrm{pre}}(\theta, \phi).&&
\end{flalign}
The pretrained parameters of the backbone and the task head are denoted by $\theta_0$ and $\phi_0$, respectively.

\subsection{Fine-tuning}

For a downstream property prediction task, we consider a labeled dataset
$\mathcal{D} = \{(x_k, y_k)\}_{k=1}^K$, where $y_k \in \mathbb{R}^d$ denotes a molecular or materials property of interest (e.g., bandgap or dielectric constant).
The dimensionality of the target property is denoted by $d \geq 1$, where $d>1$ allows for simultaneous prediction of multiple properties.
In this work, we restrict our study to the case $d=1$, i.e., only a single property is predicted at a time.
We denote the task-specific prediction head by $h_\psi$.
The resulting property prediction model is defined as
\begin{equation}
 \hat{y}_k = h_\psi\!\left( f_\theta(x_k) \right).
\end{equation}
The supervised loss for property prediction is given by
\begin{equation}\label{eq:prop-loss}
    \mathcal{L}_{\mathrm{task}}(\theta, \psi) =
    \sum_{k\in \mathcal{B}}
    \mathcal{L}_P(\hat{y}_k, y_k),
\end{equation}
where $\mathcal{B}$ denotes a mini-batch sampled from the dataset $\mathcal{D}$, and $\mathcal{L}_P$ is an error metric that quantifies the discrepancy between the predicted and reference property values.
In the following, we provide the mathematical formulations of the four training strategies demonstrated schematically in Fig.~\ref{fig:overview}.

\textit{From scratch.}
As a baseline, all model parameters $\{\theta, \psi\}$ are randomly initialized, and the property prediction model is trained by minimizing the loss in Eq.~\eqref{eq:prop-loss}:
\begin{flalign}
\mathrm{(Scratch)} && \min_{\theta, \psi}\; \mathcal{L}_{\mathrm{task}}(\theta, \psi) &&
\end{flalign}
This training strategy serves as a control experiment to quantify the actual performance gains provided by pretraining.
The performance of models trained from scratch is often limited by the size of the downstream dataset $\mathcal{D}$, as graph neural networks tend to overfit when optimized on small-scale labeled data without informative prior representations.
In practice, early stopping based on validation performance is applied to mitigate overfitting.

\textit{Linear probing (LP).}
In LP, the pretrained backbone representation is treated as a frozen featurization module, and only the parameters of the task-specific prediction head are optimized:
\begin{flalign}
\mathrm{(LP)} && 
\begin{aligned}
    &\min_{\psi}\; \mathcal{L}_{\mathrm{task}}(\theta_0, \psi),\\
    &\text{Initialization:}\quad \psi \leftarrow \mathrm{rand.},
\end{aligned} &&
\end{flalign}
where $\psi \leftarrow \mathrm{rand.}$ indicates that the task-head parameters $\psi$ are randomly initialized before training.
LP is computationally efficient and preserves the robust representations learned during pretraining.
However, freezing the backbone also limits the model’s expressiveness and its ability to adapt representations to downstream tasks that exhibit a substantial shift in prediction objectives relative to the pretraining tasks.

\textit{Fine-tuning (FT).}
In standard FT, the backbone parameters are initialized from the pretrained weights $\theta_{0}$, while the task-specific head parameters are randomly initialized.
All parameters $\{\theta, \psi\}$ are then updated end-to-end using the downstream task objective:
\begin{flalign}
\mathrm{(FT)} && 
\begin{aligned}
    &\min_{\theta,\psi}\; \mathcal{L}_{\mathrm{task}}(\theta, \psi),\\
    &\textrm{Initialization:}\quad
    \theta \leftarrow \theta_0,\;
    \psi \leftarrow \mathrm{rand.} &&
\end{aligned}
\end{flalign}
Although this optimization allows the representations to adapt to the downstream task and better fit the target labels $y_i$, it often induces representation collapse, as discussed in \cref{sec:results}.

\textit{Multi-task fine-tuning (MFT).}
MFT reintroduces the original pretraining task (energy and force prediction) as an auxiliary objective during fine-tuning.
When no prior knowledge about the downstream dataset is available, all pretraining datasets can be included in the auxiliary objective.
Alternatively, for molecular property prediction tasks, molecular pretraining datasets such as SPICE2 may be selected, while for materials property prediction tasks, materials pretraining datasets such as OMat24 can be used.
We denote the selected pretraining dataset by $D_{m^\ast}$ and define the regularization loss as
\begin{align}\nonumber
    &\mathcal L_{\mathrm{reg}} (\theta, \phi) \\
    &= 
    \sum_{k\in \mathcal{B}_{m^\ast}}
    \lambda_E \mathcal{L}_E(\hat{E}_{m^\ast k}, E_{m^\ast k})
    + \lambda_F \mathcal{L}_F(\hat{F}_{m^\ast k}, F_{m^\ast k}),
\end{align}
where $\mathcal{B}_{m^\ast}$ denotes a mini-batch of configurations sampled from dataset $D_{m^\ast}$.
The total MFT objective is then defined as
\begin{equation}
    \mathcal L_{\mathrm{MFT}} (\theta,\psi,\phi) = 
    \mathcal L_{\mathrm{task}} (\theta,\psi)
    + \lambda_{\mathrm{reg}} \mathcal L_{\mathrm{reg}}(\theta,\phi),
\end{equation}
where $\lambda_{\mathrm{reg}}$ controls the relative weight of the regularization loss with respect to the downstream task loss.

The optimization problem is given by
\begin{flalign}
\mathrm{(MFT)} && 
\begin{aligned}
    &\min_{\theta,\psi,\phi}\; \mathcal{L}_{\mathrm{MFT}}(\theta, \psi, \phi),\\
    &\mathrm{Initialization:}\quad 
    \theta \leftarrow \theta_0,\;
    \psi \leftarrow \mathrm{rand.},\;
    \phi \leftarrow \phi_0,
\end{aligned} &&
\end{flalign}
where all parameters $\{\theta,\psi,\phi\}$ are jointly optimized.
The backbone parameters $\theta$ and the force-field task head parameters $\phi$ are initialized from their pretrained values, while the downstream task head parameters $\psi$ are randomly initialized.
As demonstrated in \cref{sec:results}, this MFT scheme effectively mitigates representation collapse by preserving the chemical and configurational priors learned during pretraining.

\section*{Acknowledgments}
 The computational resources utilized in this work were provided by the AI for Science Institute. 
The work of Chengqian Zhang, Duo Zhang, Yuzhi Zhang, Guolin Ke, and Han Wang was supported by the National Key R\&D Program of China (Grant No.~2022YFA1004300).
Han Wang was additionally supported by the National Natural Science Foundation of China (Grant Nos.~12525113 and~12561160120).
Tiejun Li acknowledges the support from National Key R\&D Program of China under grant 2021YFA1003301, and National Science Foundation of China under grant 12288101.

\section*{Data availability}
The training and inference codes of multi-task fine-tuning, standard fine-tuning and linear probing on DPA3 model are available in the DeePMD-kit repository(\url{https://github.com/deepmodeling/deepmd-kit}).
The scripts used to reproduce the reported results and figures in this paper are available in \url{https://github.com/Chengqian-Zhang/Multitask-finetuning}.

{
\bibliographystyle{ieee}
\bibliography{ref}
}

\newpage

\appendix
\onecolumn

\section{Related Work}
\textit{Pretraining and fine-tuning for property predictions.}
In the pretraining stage, atomistic representations are learned from large-scale datasets and subsequently transferred to downstream tasks. 
At present, two mainstream paradigms dominate pretraining for 3D atomistic systems.
The first paradigm is self-supervised pretraining, which typically focuses on denoising-based objectives applied to equilibrium structures, where per-atom forces are close to zero~\cite{zaidi2023pretraining, zhou2023unimol, pmlrv202feng23c}. 
This line of work has recently been extended to non-equilibrium configurations by explicitly incorporating atomic force information into the learning objective~\cite{liao2024generalizing, liu2025learning}.
The second paradigm is supervised pretraining, which directly trains models on large-scale molecular or materials datasets annotated with energies, forces, and, in some cases, stress tensors~\cite{yang2024mattersim, jia2025deriva, shoghi2023molecules, kong2025mattertune}.

In the fine-tuning stage, most of the works mentioned above adopt a standard fine-tuning paradigm, in which the backbone parameters are initialized from a pretrained model and jointly optimized with an additional task-specific prediction head. 
JMP~\cite{shoghi2023molecules} further employs layer-wise learning rate decay (LLRD) \cite{howard2018universal} to alleviate overfitting. 
Nevertheless, in this work we still observe pronounced representation collapse in JMP-L, underscoring the persistent challenge of safe adaptation in downstream tasks.
RoFt-Mol \cite{liu2025roft} provides a systematic evaluation of existing fine-tuning strategies across a wide range of molecular property prediction tasks and demonstrates that the optimal fine-tuning mechanism strongly depends on both the pretraining objective and the downstream task type.
In contrast to these studies, we focus specifically on improving OOD generalization of pretrained 3D atomistic models, an underexplored yet critical problem in realistic molecular and materials design scenarios.

\textit{Representation collapse in CV and NLP.}
In computer vision (CV) and natural language processing (NLP), foundation models have demonstrated remarkable zero-shot and transfer capabilities~\cite{chen2020moco, pmlr-v139-radford21a, guo2025deepseek}.
However, a growing body of work has shown that adapting these pretrained models to specific downstream tasks often induces \textit{representation collapse}, accompanied by a pronounced degradation in OOD generalization~\cite{aghajanyan2021better, ananya2022distort, Wortsman_2022_CVPR}.
To mitigate this issue, a variety of strategies have been proposed to preserve pretrained knowledge and reduce overfitting during fine-tuning.
These include optimization-based constraints, such as using a smaller learning rate than that employed during pretraining~\cite{Kornblith_2019_CVPR, Li2020Rethinking} or adopting layer-wise learning rate schedules~\cite{Ro_Choi_2021, Shen_Liu_Qin_Savvides_Cheng_2021}, as well as architectural constraints that freeze subsets of model parameters while fine-tuning only selected components~\cite{lee2023surgical, pmlr-v162-evci22a}.
Other approaches seek to update the entire model by interpolating between pretrained and fine-tuned weights~\cite{Wortsman_2022_CVPR}, or by explicitly penalizing deviations in the latent representation manifold~\cite{li2018delta}.
In addition, incorporating auxiliary training objectives inherited from the pretraining stage has proven effective in improving fine-tuning robustness and generalization~\cite{Ge_2017_CVPR, liu2022improved, ouyang2022training}.
Our proposed MFT follows this paradigm, but grounds the auxiliary objective in physically meaningful force-field supervision, making it particularly well suited for molecular and materials property prediction.

\textit{OOD generalization in molecular and materials design.}
OOD generalization is a central yet insufficiently explored challenge in molecular and materials discovery.
Recent large-scale benchmarks demonstrate that state-of-the-art molecular property predictors suffer substantial performance degradation under OOD settings relative to ID evaluation~\cite{antoniuk2025boom, shimakawa2024extrapolative}, a trend that is likewise observed in inorganic materials prediction~\cite{sadman2024oodmatbench, li2025probing}.
Prior efforts to address this limitation have primarily focused on uncertainty quantification (UQ) to assess predictive reliability~\cite{soleimany2021evidential}, and on leveraging UQ-guided active learning to generate new training data and improve model generalization~\cite{yin2023evaluating}.
Complementary approaches incorporate large-scale unlabeled data via semi-supervised learning~\cite{wu2024instructorinspired} or adapt bilinear transduction frameworks to enhance OOD robustness in both molecular and materials domains~\cite{segal2025known, netanyahu2023learning}.
Despite these advances, OOD generalization has rarely been examined in the context of fine-tuning large pretrained  atomistic models for downstream property prediction.
In particular, the systematic degradation of OOD performance during fine-tuning, its connection to representation collapse, and principled mitigation strategies remain largely unexplored.

\section{Representation collapse of JMP-L}
\label{sec:jmpl}

\begin{figure}[h]
  \begin{center}
    \centerline{\includegraphics[width=0.5\columnwidth]{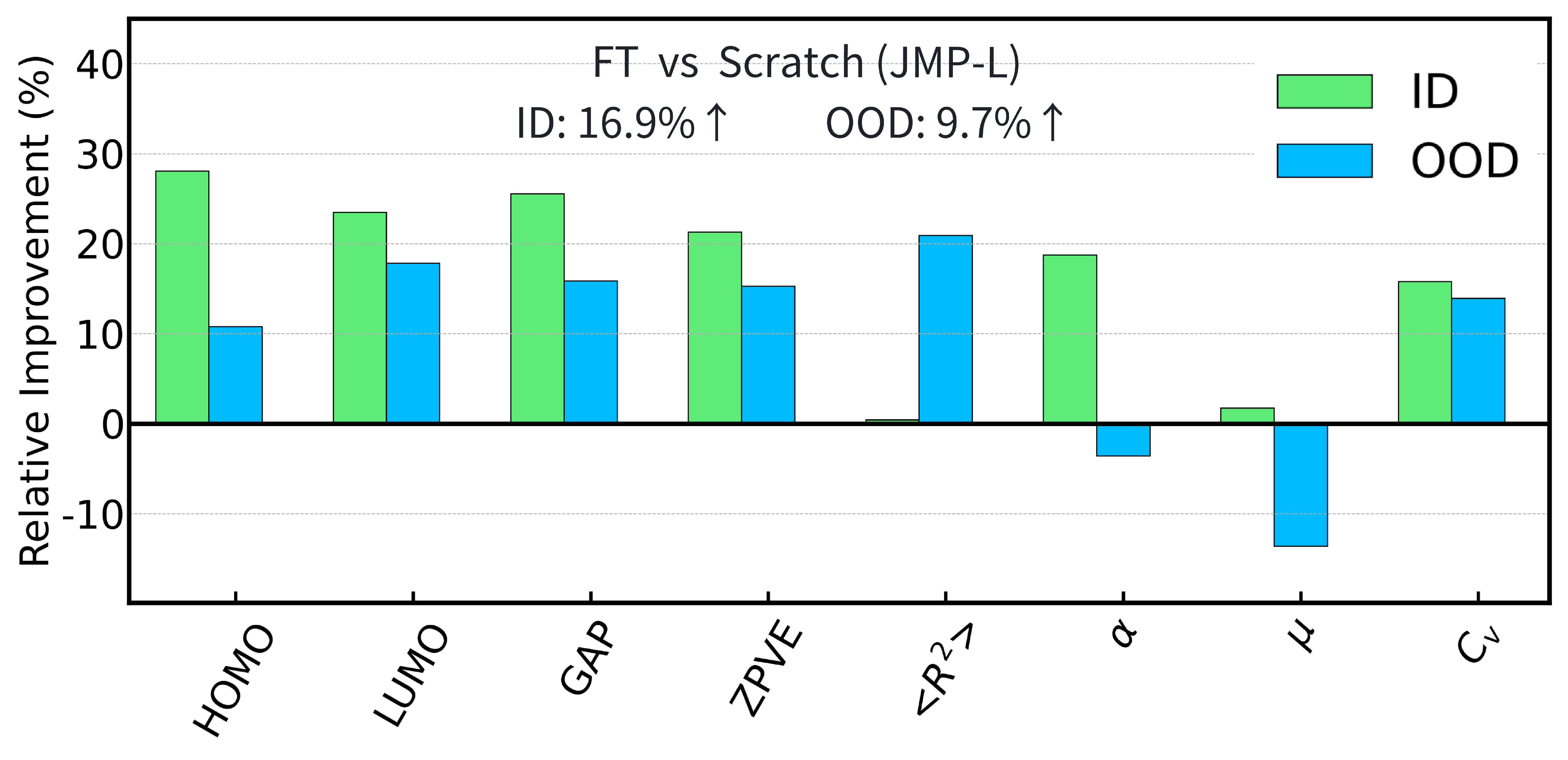}}
    \caption{Analysis of representation collapse of JMP-L. 
    Relative performance improvements of fine-tuning (FT) over the training from Scratch baseline.}
    \label{fig:jmp_ood}
  \end{center}
  \vskip -0.3in
\end{figure}

\begin{figure}[h]
  \begin{center}
    \centerline{\includegraphics[width=0.8\columnwidth]{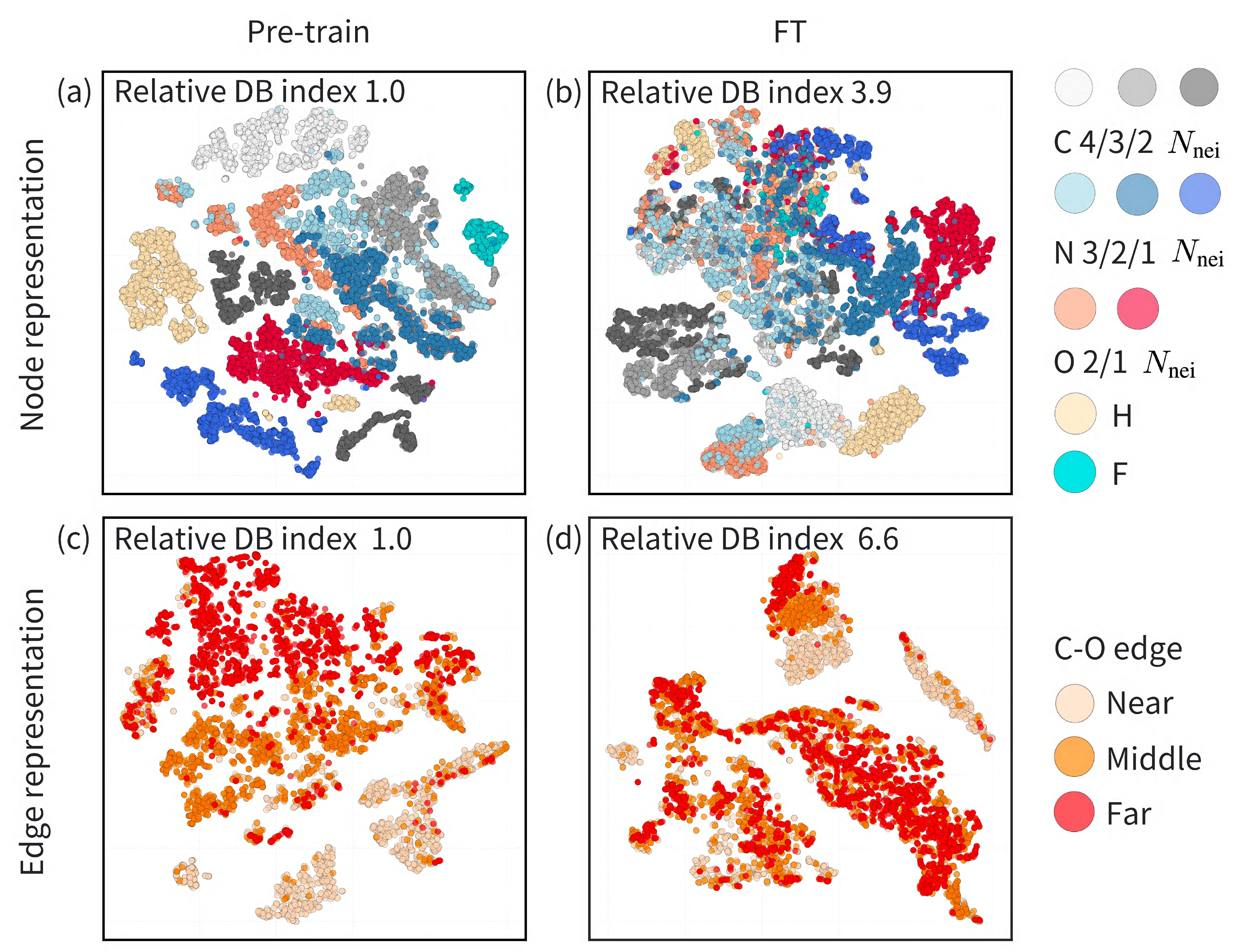}}
    \caption{t-SNE visualizations of atomic (node) and edge representations representations of the last backbone layer of JMP-L on QM9 LUMO task, reduced to 2-dimension space.
    The Davies–Bouldin (DB) index values are normalized to the pretrained model baseline.}
    \label{fig:jmp_node}
  \end{center}
  \vskip -0.3in
\end{figure}

\begin{table}[t]
  \caption{
  Quantification of distribution shift between JMP-L pretrained and fine-tuned representations using 1-Wasserstein distance ($\mathcal{W}_{\textrm{Pretrain}\rightarrow\textrm{FT}}$, $\mathcal{W}_{\textrm{Pretrain}\rightarrow\textrm{MFT}}$). 
  Metrics are reported for the QM9 LUMO dataset across training, ID, and OOD sets.
  }
  \label{tab:w1_jmp}
  \begin{center}
        \begin{tabular}{lccc}
          \toprule
            & Train & ID & OOD\\
          \midrule
          $\mathcal{W}_{\textrm{Pretrain}\rightarrow\textrm{FT}}$  & 7.54 & 7.12 & 8.58 \\
          \bottomrule
        \end{tabular}
  \end{center}
  \vskip -0.1in
\end{table}

In this section, we investigate the representation collapse when fine-tuning the pretrained JMP-L model~\cite{shoghi2023molecules}.
We use the exactly same data split of QM9 dataset as DPA-3.1-3M, as described in \cref{sec:results}.
Relative to training from scratch, fine-tuning yields an average improvement of 16.9\% in ID accuracy, but a 9.7\% improvement in OOD accuracy, as reported in ~Fig.~\ref{fig:jmp_ood}.
Across the eight property prediction tasks, the relative OOD improvement is consistently smaller than the corresponding ID improvement, with the sole exception of electronic spatial extent($\langle R^2 \rangle$), which aligns with the observation of DPA-3.1-3M as shown in Fig.~\ref{fig:results}(b).
Notably, for the isotropic polarizability($\alpha$) and the dipole moment($\mu$), the fine-tuned model underperforms the from-scratch baseline in the OOD regime by 3.6\% and 13.6\%, respectively.
Notably, the OOD improvement of JMP-L is more than DPA-3.1-3M, it is reasonable because JMP-L utilize both Layer-wise Learning Rate Decay (LLRD) strategy and a smaller learning rate than pretraining to mitigate overfitting.
However, we still observe representation collapse and OOD performance degradation phenomenon in JMP-L, showing the challenge of safe adaptation for downstream task.

Fig.~\ref{fig:jmp_node}(a) visualizes the atomic(node) representations of the last backbone layer of the pretrained JMP-L model.
The colors are following DPA-3.1-3M in Fig.~\ref{fig:results}.
As shown in Fig.~\ref{fig:jmp_node}(a), atoms of the same species and identical covalent environments tend to form clusters, thought the clustering performance is less than DPA-3.1-3M as illustrated in Fig.~\ref{fig:results}, such as some overlap between representations of oxygen and nitrogen.
After fine-tuning, however, significant overlap emerges between no matter distinct species or different local covalent environments of one species as depicted in Fig.~\ref{fig:jmp_node}(b).
Almost all distinct species with different local environments mix together, with Davies-Bouldin(DB) index increasing by a factor of 3.9 relative to the pretrained representations, indicating a substantial blurring and loss of cluster separability.

Fig.~\ref{fig:jmp_node}(c) visualizes the edge representations of carbon-oxygen(C-O) pairs, which are further stratified to near, middle and far categories according to their pairwise Euclidean distances.
As shown in Fig.~\ref{fig:jmp_node}(c), we can distinct the three types of edges in the pretrained representation space, far edges cluster at left top and near edges cluster at right bottom, though there is some mixing between middle and far edges.
However, fine-tuning induces an evident structural degradation of this latent organizations, as shown in Fig.~\ref{fig:jmp_node}(d).
The representations of middle and far edges overlap significantly.
This loss of configurational resolution is quantitatively reflected by a 6.6-fold increase in the DB index.

Furthermore, we quantify the drift between representation distributions of the pretrained and fine-tuned JMP-L models using the 1-Wasserstein distance($\mathcal{W}$).
As shown in \cref{tab:w1_jmp}, like DPA-3.1-3M, the drift distances for the training and ID test distributions are very close.
However, the drift of ID test distribution(7.12) is much larger than DPA-3.1-3M(3.75), which implies a significant representation modifications after fine-tuning.
Besides, the OOD test distribution has a larger drift than training and ID test distribution, causing overfitting. 

\section{Benchmark details}
\label{sec:benchmark_detail}

\begin{table}[h]
  \caption{Information of benchmarking properties of QM9}
  \label{tab:qm9_info}
  \centering  
  \begin{spacing}{1.1}
  \begin{tabular}{c | C{1.5cm}| C{1.5cm} | C{8cm}}
    \toprule
    Property& Samples & Units & Description \\ \hline
    $\alpha$ & 130k & $a_0^3$  &isotropic polarizability \\
    $C_v$ & 130k &$\frac{cal}{mol~K}$ & heat capacity \\
    HOMO & 130k &Hartree & highest occupied molecular orbital energy\\
    LUMO  & 130k &Hartree & lowest unoccupied molecular orbital energy\\
    GAP  & 130k &Hartree &  HOMO-LUMO gap \\
    $\mu$  & 130k &Debye &  dipole moment \\
    $\langle R^2 \rangle$& 130k &$a_0^2$ &  electronic spatial extent \\
    ZPVE  & 130k &Hartree &  zero point vibrational energy \\
    \bottomrule
  \end{tabular}
  \end{spacing}
\end{table}

\begin{table}[h]
  \caption{Information of benchmarking properties of MatBench}
  \label{tab:matbench_info}
  \centering  
  \begin{spacing}{1.1}
  \begin{tabular}{c | C{1.5cm} | C{2.5cm}| C{4cm}}
    \toprule
    Property & Samples & Units & Description \\ \hline
    dielectric  & 4764 & Unitless & Refractive index\\
    elasticity  & 10987 & log10(GPa) & Shear modulus\\
    perovskites  & 18928 & eV/unit cell & Formation energy\\
    \bottomrule
  \end{tabular}
  \end{spacing}
\end{table}

\subsection{OOD data split details of QM9}
\label{sec:qm9_detail}
For the model performance evaluation for small molecular dataset QM9~\cite{ramakrishnan2014quantum}, we use the same data split as BOOM~\cite{antoniuk2025boom}.
Training, ID test and OOD test splits are generated based on the underlying property distributions of the datasets.
For each of the eight molecular properties in \cref{tab:qm9_info}, OOD splits are constructed by fitting a Kernel Density Estimator (KDE) with a Gaussian kernel to the property values to determine the probability density of each molecule. 
Molecules with the lowest 10\% of probability scores are assigned to the OOD split for the QM9 dataset, which effectively isolates samples at the tails of the distribution for typical molecular property distributions.
Unlike traditional partitioning via fixed cut-off values, this density-based approach allows for the capture of low-probability samples within general distributions that may be non-unimodal or complex.
The ID test sets are then generated by randomly sampling 10\% sample from the remaining molecules, while all residual data is reserved for training.

\subsection{OOD data split details of MatBench}
\label{sec:matbench_detail}

The following data splitting strategies on MatBench~\cite{dunn2020benchmarking}, originally developed by Ref.~\cite{sadman2024oodmatbench}, were employed to evaluate model performance on inorganic materials across different OOD regions in the feature and property space. 
These methods rely on the Orbital-Field Matrix (OFM)~\cite{pham2017machine}, which represents materials by encoding electronic structures and orbital interactions to provide a physically interpretable feature space.

\textit{Leave-one-cluster-out(LOCO)}. 
In this approach, the dataset is partitioned into 50 clusters using the $k$-means algorithm based on OFM features.
Model evaluation is conducted iteratively, where each cluster serves as a test set while the remaining clusters are used for training and validation. 
This ensures that the model is tested on groups of materials that are structurally distinct from the training data.

\textit{Single-point targets with the lowest structure density (SparseXsingle)}. 
This method identifies structurally unique materials by projecting the 1024-dimensional OFM features into a 2D space using t-distributed Stochastic Neighbor Embedding (t-SNE). 
A Gaussian Kernel Density Estimation (KDE) is applied to this 2D manifold to identify the 500 samples with the lowest density. 
These samples are grouped into 50 clusters via $k$-means, and one representative sample is extracted from each cluster to form a 50-point test set.

\textit{Single-point targets with the lowest property density (SparseYsingle)}. 
Focusing on outliers in the property distribution, this method estimates the density of the target property values ($y$-values) using Gaussian Kernel Density Estimation. 
The 500 samples exhibiting the lowest property density are selected and partitioned into 50 clusters through $k$-means. 
One sample is then selected from each cluster, yielding a test set that represents rare or extreme property regimes.

\textit{Cluster targets with the lowest structure density (SparseXcluster)}.
This method extends the SparseXsingle logic by forming target clusters rather than individual points. 
After the 50 representative sparse samples are identified, the $N$ nearest neighbors for each sample are included in the test set. 
These neighbors are determined by the Euclidean distance in the OFM feature space, with a constraint ensuring that no sample is assigned to multiple clusters.

\textit{Cluster targets with the lowest property density (SparseYcluster).}
Similarly, this method builds upon the SparseYsingle approach. 
Once the 50 samples from low-density property regions are identified, the test set is expanded to include their $N$ nearest neighbors in the OFM feature space. 
This allows for the evaluation of model performance on localized regions of the material space associated with rare property values.

Furthermore, the 5-fold random split used in \cref{table:material} are sourced from Ref.~\cite{dunn2020benchmarking}.

\subsection{Detail results of MatBench}
\begin{figure}[h]
  \begin{center}
    \centerline{\includegraphics[width=0.8\columnwidth]{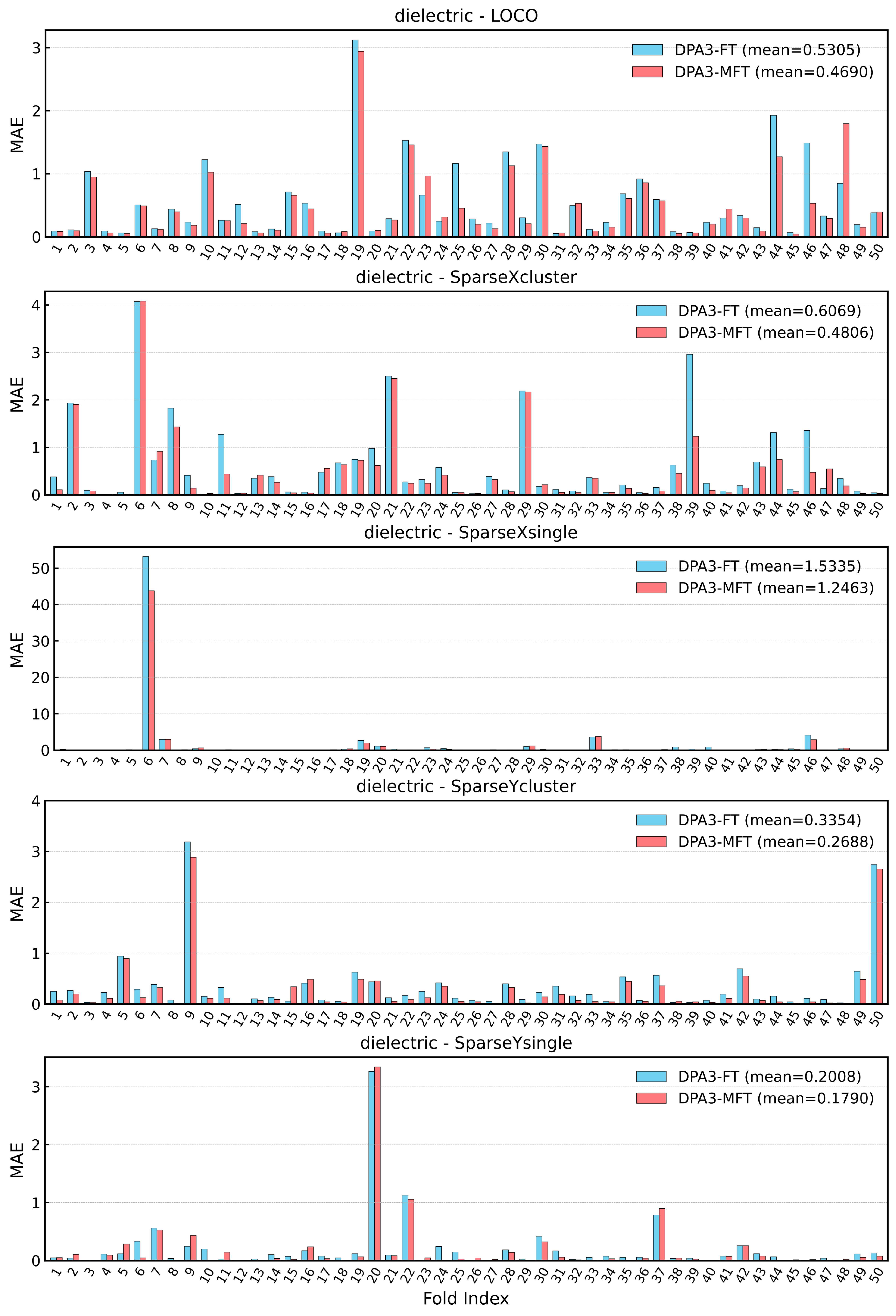}}
    \caption{Distribution of the MAEs for each fold of DPA3-FT and DPA3-MFT on the dielectric dataset across five different OOD splits.}
    \label{fig:dielectric}
  \end{center}
  \vskip -0.3in
\end{figure}

\begin{figure}[h]
  \begin{center}
    \centerline{\includegraphics[width=0.8\columnwidth]{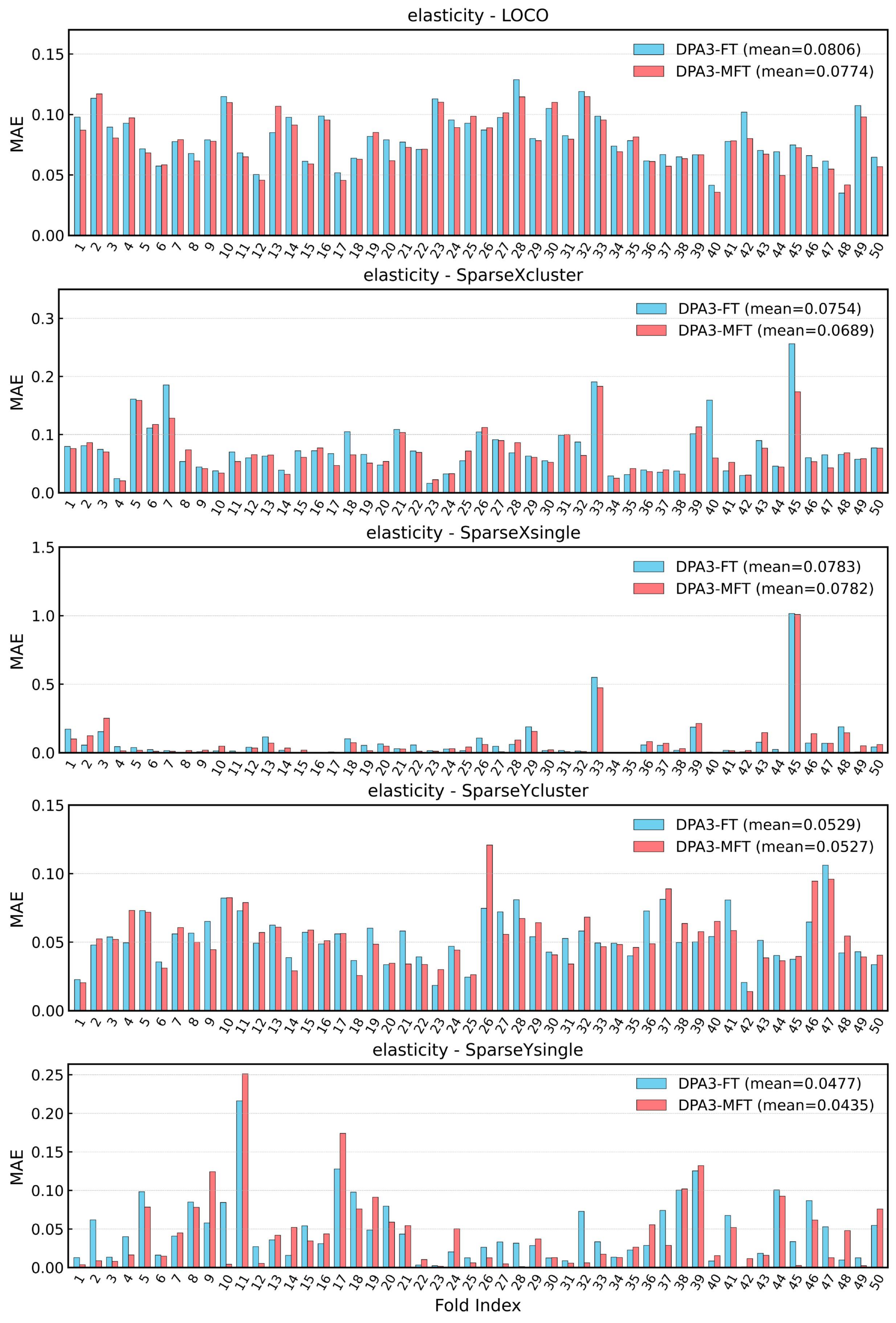}}
    \caption{Distribution of the MAEs for each fold of DPA3-FT and DPA3-MFT on the elasticity dataset across five different OOD splits.}
    \label{fig:elasticity}
  \end{center}
  \vskip -0.3in
\end{figure}

\begin{figure}[h]
  \begin{center}
    \centerline{\includegraphics[width=0.8\columnwidth]{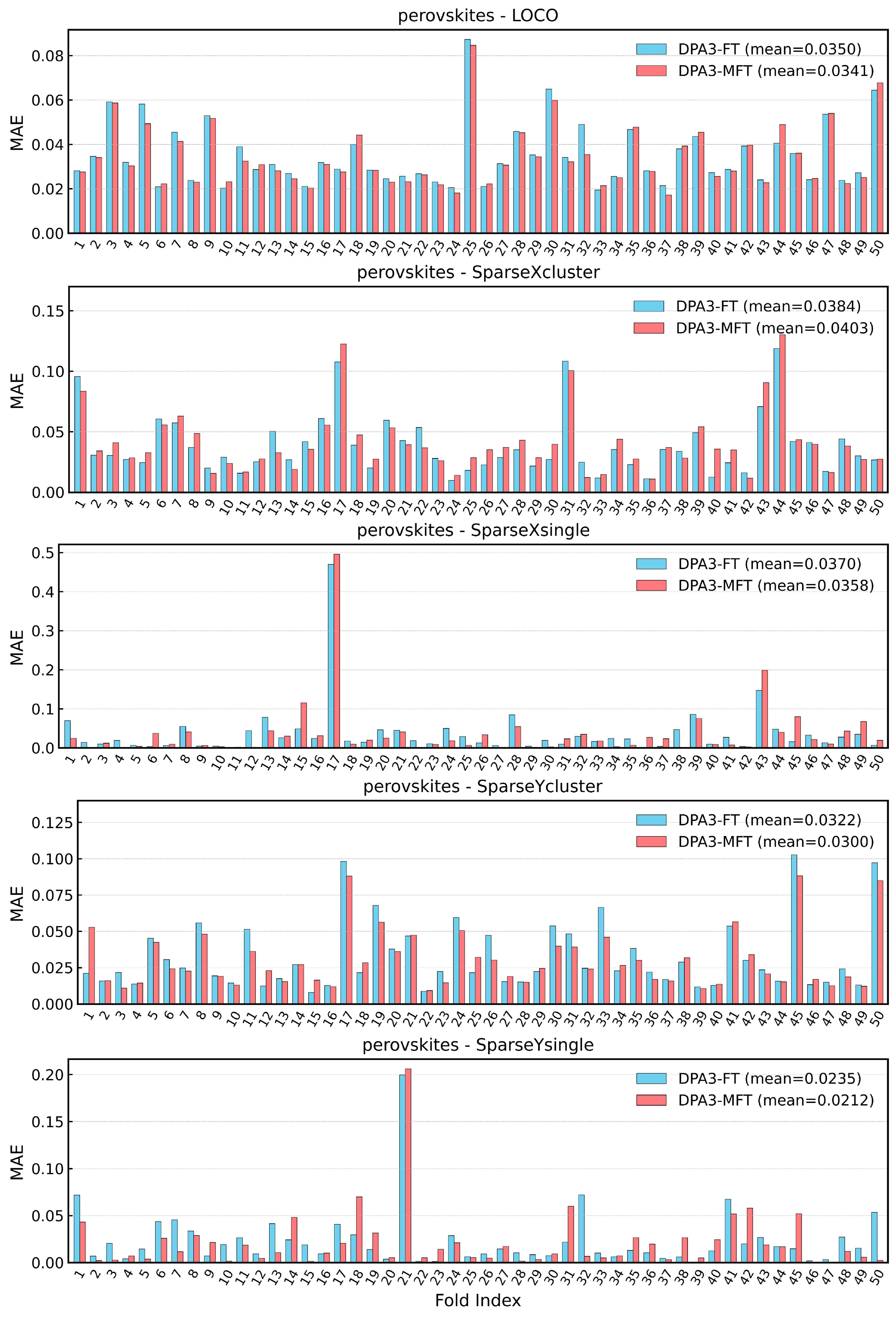}}
    \caption{Distribution of the MAEs for each fold of DPA3-FT and DPA3-MFT on the perovskites dataset across five different OOD splits.}
    \label{fig:perovskites}
  \end{center}
  \vskip -0.3in
\end{figure}

The fold-wise MAE plots for DPA3-FT and DPA3-MFT on the dielectric dataset, elasticity dataset, and perovskites dataset are presented in Fig.~\ref{fig:dielectric}, Fig.~\ref{fig:elasticity} and Fig.~\ref{fig:perovskites}, respectively.
The distribution of the MAE for 50 folds show that only a small subset of challenging folds are responsible for the overall MAE for both DPA3-FT and DPA3-MFT, which aligns with previously reported observations for baseline models in Ref.~\cite{sadman2024oodmatbench}.
Despite these shared distributional characteristics, DPA3-MFT consistently outperforms both DPA3-FT and baseline models as shown in \cref{table:material}.

\section{Implementation details}
\subsection{Representation analysis details}
\label{sec:rep_detail}
In Fig.~\ref{fig:results}(d-f) visualize the atomic(node) representations of the pretrained and fine-tuned model.
Atoms of different chemical species are indicated by distinct colors, while atoms of the same species are further differentiated by their number of covalent neighbors.
Here, we consider two atoms with a distance less than 1.15 times the sum of covalent radii as neighbors to each other.
We randomly sample 2000 atomic representations for each type of local covalent environment of each atom species in OOD test set for next-step analysis.
Subsequently, we apply the t-distributed stochastic neighbor embedding(t-SNE)~\cite{JMLR:v9:vandermaaten08a} for dimension reduction converting the 128-dimension atomic representations to 2-dimension for visualization via scikit-learn package~\cite{scikit-learn}.
We calculate Davies–Bouldin(DB)~\cite{dbindex1979} index in 2-dimension representation space to quantitatively assess the clarity and separation of the representation clusters via scikit-learn package.
The definition of DB index is
\begin{equation}
    \text{DB index} = \frac{1}{N}\sum_{i=1}^N \max_{i \neq j}(\frac{S_i + S_j}{M_{ij}})
\end{equation}
where $M_{ij}$ is the distance between vectors which are chosen as characteristic of clusters $i$ and $j$, and $S_i$ and $S_j$ are the dispersions of clusters $i$ and $j$, respectively.
Specifically, $S_i$ and $M_{ij}$ are defined as
\begin{equation}
    S_i = \{\frac{1}{T_i}\sum_{j=1}^{T_i}|X_j - A_i|^q\}^{\frac{1}{q}}~~~~M_{ij}=\{\sum_{k=1}^n|a_{ki} - a_{kj}|^p\}^{\frac{1}{p}}
\end{equation}
where $T_i$ is the number of vectors in cluster $i$, $A_i$ is the centroid of cluster $i$, $a_{ki}$ is the $k$th component of the $n$-dimension vector $a_i$, which is the centroid of cluster $i$.
Here we define $p=2$ and $q=1$, which means $M_{ij}$ is the Euclidean distance between centroids, $S_i$ becomes the average Euclidean distance of vectors in cluster $i$ to the centroid of cluster $i$.
Lower DB index values indicate more distinct and cohesive clusters.

Both DPA-3.1-3M and JMP-L employ edge representations to encode pairwise atomic relationships within a preset cutoff radius.
We visualize the edge representations of carbon-hydrogen(C-H) and carbon-oxygen(C-O) pairs, two prevalent edge types in the downstream tasks for DPA-3.1-3M, and visualize C-O edge representations for JMP-L.
These edges are further stratified into \emph{near}, \emph{middle}, and \emph{far} categories according to their pairwise Euclidean distances. 
Here, we consider the edge where the distance between atomic pairs is less than 1.15 times the sum of the covalent radii of two atoms to be the \emph{near} edge, i.e. the atom pairs form covalent bonds.
We consider the edge where the distance between atomic pairs between 3 angstroms and 4 angstroms is the \emph{middle} edge.
The atom pairs in \emph{middle} edges are nonbonded but their features are updated with angular information with angular cutoff in DPA-3.1-3M being 4 angstroms.
We consider the edge where the distance between 5 angstroms and 6 angstroms is the \emph{far} edge, as the cutoff of DPA-3.1-3M is 6 angstroms.
Compared to \emph{middle} edges, \emph{far} edges are also nonbonded but the atom pairs' features are updated without angular information.
Similar to the analysis to atomic representation, we randomly sample 2000 edge representations for each type of edge in OOD test set, then use t-SNE to convert high-dimension edge representations to 2-dimension space for visualization and use DB index to assess the clarity and separation of edge representation clusters.

We formally quantify the similarity, or equivalently the drift, between atomic representation distributions of the pretrained and fine-tuned models using the 1-Wasserstein distance($\mathcal{W}_1$), briefly denoted by $\mathcal{W}$ in this paper.
In practice, we sample 5000 atomic representations of the pretrained and fine-tuned models respectively.
The discrete optimal transport problem is defined as:
\begin{equation}
    \mathcal{W} = \min_{\gamma}\sum_{i,j}\gamma_{i,j}C_{i,j}
\end{equation}
where $\gamma \in \mathbb{R}^{5000\times 5000}$ is the optimal transport matrix, $C \in \mathbb{R}^{5000\times 5000}$ is cost matrix with $C_{i,j}$ defined as the Euclidean distance between the $i$-th pretrained representation and the $j$-th fine-tuned representation in the origin 128-dimension space.
The results are computed via the POT package \cite{flamary2021pot}.

\subsection{Model architecture details}
Following the notation in the main text, we denote the GNN backbone used to learn atomistic representations as $f_\theta$, with trainable parameters $\theta$, while denote the force-field task head and downstream property prediction head as $g_\phi$ and $h_\psi$, with trainable parameters $\phi$ and $\psi$, respectively.
The architecture of $f_\theta$ and $g_\phi$ is exactly the same as DPA-3.1-3M~\cite{zhang2025graphneuralnetworkera} and we will not elaborate details in this article.

For a downstream property prediction task, we consider a labeled training dataset
$\mathcal{D} = \{(x_k, y_k)\}_{k=1}^K$, where each data point consists of a three-dimensional atomic structure $x_{k}$, and $y_k \in \mathbb{R}^d$ denotes a molecular or materials property of interest (e.g., bandgap or dielectric constant).
Formally, each structure is represented as $x_{k} = (R_{k}, Z_{k})$, where $R_{k} \in \mathbb{R}^{N_{k} \times 3}$ denotes the Cartesian coordinates of the $N_{k}$ atoms, and $Z_{k} \in \mathbb{N}^{N_{k}}$ denotes their atomic numbers.
The dimensionality of the target property is denoted by $d \geq 1$, where $d>1$ allows for simultaneous prediction of multiple properties.
In this work, we restrict our study to the case $d=1$, i.e., only a single property is predicted at a time.
Generally, the formulation of $h_\psi$ is
\begin{equation}
     \label{eq:h_psi}
     \hat{y}_k = h_\psi\!\left( f_\theta(x_k) \right) = \begin{cases}
     \sum_{i=1}^{N_k} [\textrm{MLP}\!\left( f_\theta(x_k)_i \right) + p(Z_{ki})], & \text{if}~y_k~\text{is extensive}.\\
     \frac{1}{N_k} \sum_{i=1}^{N_k} [\textrm{MLP}\!\left( f_\theta(x_k)_i \right) + p(Z_{ki})], & \text{if}~y_k~\text{is intensive}.
     \end{cases}
\end{equation}
where $\textrm{MLP}$ refers to Multilayer Perceptron, which is employed in this study with a specific configuration of 3 layers and 240 neurons per layer.
Furthermore, $p(Z_{ki})$ is a constant bias to the atomic property contribution according to the atomic number $Z_{ki}$.
This property bias are determined by a least-square fitting of the properties in the training data.
More precisely, we have $K$ training data frames, and within the $k$-th frame, we have $N_{kz}$ atoms with atom number $z$, and the labeled property of the frame is denoted by $y_{k}$.
Then a linear system is solved in the least-square sense.

\begin{equation}
    \sum_z N_{kz}p(z) = \begin{cases}
        y_{k}, &\text{if}~y_k~\text{is extensive}\\
        N_k y_{k},&\text{if}~y_k~\text{is intensive}
    \end{cases}~~~k=1,...,K
\end{equation}

We regard HOMO, LUMO, GAP in QM9 and dielectric, elasticity, perovskites in MatBench as intensive properties, while regard QM9 $\alpha$, ZPVE and $C_v$ as extensive properties.
There are three special cases: dielectric, $\langle R^2 \rangle$ and $\mu$.
As to dielectric, we do not use the property bias term $p(Z_{ki})$ in \cref{eq:h_psi}, because there are element species in the test set that did not appear in the training set.
As to QM9 $\langle R^2 \rangle$ and $\mu$, we use customized task head architecture to better utilize priors.
Specifically, we use the same prediction head formulation as previous work~\cite{tholke2021equivariant, shoghi2023molecules, aykent2025gotennet} for $\langle R^2 \rangle$ prediction and the norm of dipole moment vectors~\cite{Zhang_PhysRevB_2020_v102_p41121} for predictions of dipole moment scalars $\mu$.

\subsection{Model hyperparameters}
The hyperparameters of training from scratch, fine-tuning(FT) and multi-task fine-tuning(MFT) utilized for DPA3 model are illustrated in \cref{table:hyperparam}.
DPA3 uses dynamic batch size based on the number of atoms $N$ in each specific system.
Formally, $\textrm{Auto:}X = \lceil \frac{X}{N} \rceil$ where $N$ denotes the number of systems in each system and $\lceil \cdot \rceil$ denotes the ceiling function which rounds the number up to the nearest integer.
Notation ``$\rightarrow$" appears in energy, force and virial loss prefactors indicate that the prefactors are changed in synchronization with the learning rate.
Besides, DPA3 employs a customized activation function termed SiLUT(SiLU threshold with Tanh) to improve numerical stability during training, which is defined as:
\begin{align}
    \textrm{SiLUT:t}(x) = 
    \begin{cases}
    \textrm{SiLU}(x), & \text{if } x \leq t.  \\ 
    \textrm{Tanh}(a \cdot(x-t)) + b, & \text{if } x > t.
\end{cases}
\end{align}
where $t$ represents the threshold, indicating when to transition from SiLU to Tanh.
Constants $a$ and $b$ are determined based on $t$ to ensure the first and second continuity of the activation function at the threshold.

Regarding the JMP model hyperparameters, we employed a batch size of 28 and conducted training over 24 epochs. 
All remaining hyperparameters were maintained in accordance with Ref.~\cite{shoghi2023molecules} for both training from scratch and fine-tuning.

\subsection{Details of data efficiency experiments}
\label{sec: data_effi_detail}
The data efficiency of MFT is evaluated through a series of experiments on the QM9 band gap (GAP) and MatBench refractive index (dielectric) prediction tasks, as depicted in Fig.~\ref{fig:data_effi_and_ablation}(a) and Fig.~\ref{fig:data_effi_and_ablation}(b). 
In these experiments, the training set size is varied by randomly down-sampling the original data across three orders of magnitude. 
This approach tests model performance in extreme low-data regimes, reaching as few as 114 data points for the QM9 GAP task and 14 data points for the MatBench dielectric task.

For the QM9 GAP prediction task, we generate 3 independent training subsets for every data scale using different random seeds and 3 independent runs are performed to ensure statistical reliability. 
The results are reported using the root-mean-square errors (RMSEs) for both ID and OOD test sets, plotted as a function of the number of training data.

In the MatBench dielectric prediction task, we utilize random splits to evaluate ID performance and leave-one-cluster-out (LOCO) splits for OOD evaluation across varying training data scales.
Unlike QM9 GAP which has one fold, MatBench dielectric employs multiple folds, specifically 5 folds for random splits and 50 folds for LOCO.
In practice, we selected the specific fold whose mean absolute error (MAE) of test sets most closely approximated the mean MAE across all folds during full-dataset FT. 
This representative fold is used for subsequent data efficiency experiments.

For the random split experiments on the dielectric task, we maintain the same sampling strategy used for QM9 GAP, utilizing 3 independent training subsets and 3 runs for every data scale. 
However, to mitigate the higher variance inherent in the LOCO split for the dielectric task, we implement a more rigorous sampling strategy. 
Specifically, for data scales ranging from $1$ down to $1/16$, we conducted 6 independent runs using 6 different random subsets.
For the smaller data scales, including $1/32$, $1/64$, $1/128$, and $1/256$, the number of trials is increased to 10 independent runs with 10 distinct random subsets. 
The resulting MAEs are plotted against the training data size to demonstrate the model's data efficiency.

\begin{table}[t]
\caption{Hyper-parameters for DPA3 models trained on QM9 and MatBench.}
\label{table:hyperparam}
\centering
\begin{threeparttable}
\begin{spacing}{1.2}
 \begin{tabular}{c|l|cc}
  \toprule
    & Hyper-parameters & QM9 & MatBench\\\hline 
   \multirow{4}{*}{General} & Optimizer & Adam & Adam \\
   & Learning rate scheduling & Exp & Exp \\
   &Maximum learning rate & 1e-3 & 1e-3 \\
   &Minimum learning rate & 1e-5 & 1e-5 \\ 
   \hline
   \multirow{5}{*}{Scratch/FT} &Number of training steps & 0.1M & 0.1M \\
   &Loss function & MSE & Smooth MAE \\ 
   & Batch size per GPU & Auto:512 & Auto:512 \\
   & Number of GPUs & 1 & 1 \\
   & Activation function & Tanh & Tanh \\\hline
   \multirow{11}{*}{MFT}
   & Number of training steps & 0.1M & 0.2M \\
   & Force-field task head & SPICE2 & OMat24 \\
   & Energy loss prefactor & 0.2 $\rightarrow$ 20 & 0.2 $\rightarrow$ 20\\
   & Force loss prefactor & 100 $\rightarrow$ 60 & 100 $\rightarrow$ 60\\
   & Virial loss prefactor & - & 0.02 $\rightarrow$ 1 \\
   & Property loss prefactor & 1 & 1 \\
   & Batch size per GPU~(property task) & Auto:512 & Auto:512 \\
   & Batch size per GPU~(force-field task) & Auto:128 & Auto:128 \\
   & Activation function~(property head) & Tanh & Tanh \\
   & Activation function~(force-field head) & SiLUT:3.0 & SiLUT:3.0\\
   & Number of GPUs & 2 & 1 \\
   \toprule
 \end{tabular}
 \end{spacing}
 \end{threeparttable}
\end{table}

\end{document}